\documentclass[11pt,aps,prd,onecolumn,groupedaddress,
superscriptaddress,preprintnumbers,notitlepage,nofootinbib]{revtex4-2}
\usepackage[utf8]{inputenc}
\usepackage{natbib} 
\usepackage{graphicx,amsthm} 
\usepackage{dcolumn} 
\usepackage{bm} 

\usepackage{amsfonts}
\usepackage[caption=false]{subfig}
\usepackage{slashed}
\usepackage{float}
\usepackage{longtable}
\usepackage{ragged2e}
\usepackage[usenames, dvipsnames]{color}
\usepackage{amssymb}
\usepackage[colorlinks = true,
            linkcolor = blue,
            urlcolor  = blue,
            citecolor = blue,
            anchorcolor = blue]{hyperref}
\usepackage{pgf}
\usepackage{titlesec}
\usepackage{xcolor}
\usepackage{tcolorbox}
\usepackage{soul}
\renewcommand\thesection{\arabic{section}.}
\renewcommand\thesubsection{\thesection.\arabic{subsection}}
\renewcommand\thesubsubsection{\thesubsection.\arabic{subsubsection}}
\titleformat{\section}[hang]{\normalfont\large\bfseries\raggedright}{\thesection}{0.3em}{}
\titleformat{\subsection}[hang]{\normalfont\normalsize\bfseries\raggedright}{\thesubsection}{0.3em}{}
\titleformat{\subsubsection}[hang]{\normalfont\normalsize\bfseries\raggedright}{\thesubsubsection}{0.3em}{}

\usepackage{fancyhdr}
\everymath{\displaystyle} 
\usepackage{newtxtext,newtxmath}

\usepackage{tikz}
\usepackage{colortbl}
\usepackage{fancybox}
\usepackage{epstopdf}
\usepackage{latexsym}
\usepackage{amssymb}
\usepackage{orcidlink}
\begin{document}
\setlength{\parskip}{0.4em}
\title{Einstein-Gauss-Bonnet quintessential inflation: From super-inflation to emergent warm inflation}

\author{Omar Chahboun \orcidlink{0000-0001-6567-846X}}
\email[]{chahboun.omar1@gmail.com}
\affiliation{\footnotesize \textit{Laboratory of High Energy Physics: Modeling and Simulation (LPHE-MS), Science Faculty,
Mohammed V University in Rabat, 4 Ibn Batouta Avenue BP 1014 RP, 10090, Rabat, Morocco}}\affiliation{\footnotesize \textit{Centre of Physics and Mathematics (CPM), 4 Ibn Batouta Avenue BP 1014 RP, 10090, Rabat,
Morocco}}

\author{Rachid Ahl Laamara \orcidlink{0000-0002-8254-9085}}

\affiliation{\footnotesize \textit{Laboratory of High Energy Physics: Modeling and Simulation (LPHE-MS), Science Faculty,
Mohammed V University in Rabat, 4 Ibn Batouta Avenue BP 1014 RP, 10090, Rabat, Morocco}}\affiliation{\footnotesize \textit{Centre of Physics and Mathematics (CPM), 4 Ibn Batouta Avenue BP 1014 RP, 10090, Rabat,
Morocco}}

\author{El Hassan Saidi \orcidlink{0000-0003-4039-6700}}

\affiliation{\footnotesize \textit{Laboratory of High Energy Physics: Modeling and Simulation (LPHE-MS), Science Faculty,
Mohammed V University in Rabat, 4 Ibn Batouta Avenue BP 1014 RP, 10090, Rabat, Morocco}}\affiliation{\footnotesize \textit{Centre of Physics and Mathematics (CPM), 4 Ibn Batouta Avenue BP 1014 RP, 10090, Rabat,
Morocco}}

\date{\today}

\begin{abstract}
\fontsize{10pt}{9.5pt}\selectfont
\begin{description}
\renewcommand\makelabel[1]{\textbf{#1}}
\item
{\normalfont\normalsize\bfseries Abstract}
\\This work presents a unified inflationary and late-time cosmology, developed within Einstein–Gauss–Bonnet gravity coupled to three scalar fields, governed by a single potential and coupling function throughout cosmic history. Permitting the Gauss–Bonnet coupling to depend on the field content, rather than assigning it arbitrarily, produces a two-phase inflationary scenario: the higher-curvature term initially drives expansion and later becomes a subdominant. The initial super-inflationary stage ends dynamically when the geometrically induced stabilization of the waterfall field fails, at a field value determined by the combination $(n-1)Q$. Coherent oscillations of the waterfall condensate serve as the initial radiation source, replacing conventional reheating and establishing the thermal bath necessary for the subsequent warm phase. Utilizing a two-stage dissipation mechanism with $\Gamma_\phi\propto T^3$, the scalar and tensor spectra are derived while retaining Gauss–Bonnet contributions, demonstrating that the growth of inflaton fluctuations is regulated by the shear viscosity of the bath, which is essential for model viability. A quartic potential, previously excluded in the cold scenario, becomes viable, with $n_s$ within the Planck $1\sigma$ band and $r$ suppressed. Agreement with the spectral index is generic, while internal consistency confines the model to a narrow band in $(n-1)Q$. Finally, the release of the third scalar near matter–radiation equality transforms a very small mass scale into a very large coupling by a logic inverse to the seesaw mechanism. At late times, the inflaton persists as dark energy.

\item {\normalfont\normalsize\bfseries Keywords:} Einstein-Gauss-Bonnet gravity -- super-inflation -- waterfall phase transition -- warm inflation -- quintessence dark energy
 
\end{description}
\end{abstract}

\maketitle

\section{Introduction}
\label{intro}

Modern cosmology has made significant advances in characterizing large-scale structure and evolution of the Universe; however, the resulting picture remains enigmatic. Precision measurements of the cosmic microwave background, galaxy clustering, and supernova distances have established the $\Lambda$CDM model as the prevailing paradigm \cite{Planck2018VI,PhysRevD.103.083533}. Notably, this model attributes about 95\% of the Universe's present energy content to two components whose physical nature remains unknown. Dark matter is inferred solely from its gravitational effects, such as the flatness of galactic rotation curves, the growth of structure, and the acoustic peaks in the microwave background \cite{1980ApJ...238..471R,RevModPhys.90.045002}. In contrast, dark energy is characterized only by an equation of state near $-1$ and a magnitude that is approximately one hundred and twenty orders of magnitude below its natural theoretical estimate \cite{RevModPhys.61.1,2006IJMPD..15.1753C}. At the opposite extreme of scale, the study of black holes probes gravity in a regime where the classical description is expected to break down, and the longstanding tensions among general relativity, quantum theory, and thermodynamics have persisted for half a century \cite{Hawking:1975vcx,2013JHEP...02..062A}. The simultaneous incompleteness of general relativity at both the largest and smallest scales suggests that it should be regarded as the low-energy limit of a more fundamental theory. This motivates the question of what corrections to general relativity might exist and where they would first become observable.

The primordial Universe provides a unique context for addressing these questions, as it probes energy scales far beyond those accessible in terrestrial experiments and leaves observable imprints. Cosmological inflation \cite{sym15091701}, originally proposed to resolve the horizon, flatness, and monopole problems of the hot big bang \cite{PhysRevD.23.347,LINDE1982389,PhysRevLett.48.1220,STAROBINSKY198099}, also achieves an unanticipated outcome, it transforms quantum fluctuations of a field into the classical density perturbations that seed all subsequent structure \cite{Mukhanov:1981xt,HAWKING1982295,PhysRevLett.49.1110,STAROBINSKY1982175,PhysRevD.28.679}. This linkage between microscopic physics and cosmological observables renders inflation empirically testable. In its simplest form, a scalar field known as the inflaton evolves slowly along a nearly flat potential, driving a quasi-de Sitter expansion and producing an almost scale-invariant spectrum of curvature perturbations \cite{LINDE1983177,LYTH19991}. Deviations from exact scale invariance are small but measurable, and are typically characterized by the scalar spectral index $n_s$ and the tensor-to-scalar ratio $r$, both of which are now sufficiently constrained to exclude broad classes of models \cite{Planck2018X,PhysRevLett.127.151301}. Notably, the simple quartic potential is excluded, as its cold-inflation prediction $r\simeq16/N_*$ exceeds observational bounds by more than an order of magnitude.

If general relativity is regarded as an effective theory, the inflationary epoch is precisely where its corrections are expected to be significant, motivating extensive research on inflation within modified gravity frameworks \cite{2010LRR....13....3D,NOJIRI20171,2007IJGMM..04..115N,2011PhR...505...59N,Clifton:2011jh}. Among the possible extensions, the Gauss-Bonnet invariant holds a distinguished status\cite{PhysRevD.77.023526,2008JCAP...09..019S}. It is the only quadratic curvature combination whose variation yields second-order field equations, ensuring that theories incorporating it are free from the Ostrogradsky instability \cite{Lovelock:1971yv}. Furthermore, it emerges naturally in the low-energy effective action of heterotic string theory \cite{Gross:1986mw,Metsaev:1987zx}. In four dimensions, the Gauss--Bonnet term is topological and does not contribute dynamically unless coupled to a scalar field, which introduces rich phenomenology; the coupling alters the propagation of tensor modes \cite{2026arXiv260809783O}, breaks the standard consistency relation between $r$ and the tensor tilt, and can suppress the tensor amplitude enough to render previously excluded potentials viable \cite{Guo:2010jr,PhysRevD.88.123508,Koh:2014bka,2021NuPhB.96315299O,2020AnPhy.42068250O,2020NuPhB.95815135O,2020PDU....3000718O,2023PhRvD.107h4007O,2025arXiv250608193O,2026PDU....5202348O}. Recent developments in four-dimensional Einstein-Gauss-Bonnet gravity as a modification of general relativity have further intensified interest in these questions \cite{PhysRevLett.124.081301,Fernandes_2022}.

A second significant modification to the canonical framework concerns inflation's termination. In the standard scenario, the inflaton is assumed to decouple during the accelerated expansion, decaying only afterward to reheat the Universe in a largely unconstrained subsequent epoch. Warm inflation \cite{PhysRevLett.74.1912,PhysRevLett.75.3218} challenges this assumption by noting that, since the inflaton must ultimately transfer its energy to the Standard Model, there is no fundamental reason to prohibit this transfer during inflation itself. Permitting such energy transfer introduces a dissipative friction term $\Gamma_\phi\dot\phi$ into the inflaton's equation of motion and, by energy conservation, a corresponding source term for radiation. As a result, a thermal bath is maintained throughout inflation rather than being generated only at its conclusion. This leads to several important consequences: the slow-roll conditions are relaxed by a factor of $1+\Gamma_\phi/3H$, perturbations are seeded by thermal rather than vacuum fluctuations \cite{BERERA2000666,Hall:2003zp}, and the Universe transitions directly into radiation domination without a distinct reheating phase \cite{Berera:2008ar}. A primary challenge is that a coupling strong enough to sustain the thermal bath often disrupts the potential's flatness through thermal corrections. This issue has been addressed by coupling the inflaton exclusively to heavy mediators that remain thermally unpopulated \cite{BasteroGil:2012cm,BasteroGil:2016qru}.

A third line of development examines the inflaton's fate after inflation ends. In the canonical framework, the field settles into a minimum and becomes irrelevant, so an independent component is needed to explain late-time acceleration. Quintessential inflation \cite{Spokoiny:1993kt,Peebles:1998qn} addresses this issue by employing a potential without a minimum. In this scenario, the field traverses the inflationary region and then evolves along a steep tail, with its energy density decreasing more rapidly than that of radiation. As a result, it remains subdominant during the radiation- and matter-dominated eras, but eventually dominates at late times, driving the observed acceleration. This approach allows a single scalar field to account for both epochs of accelerated expansion, and the coincidence between the dark energy scale and the present Hubble rate arises from the field's evolution rather than from independent tuning \cite{galaxies9040073}. The concept has been realized in various settings: for exponential potentials, where late-time behavior follows scaling or tracker trajectories that are insensitive to initial conditions \cite{2021EPJC...81..105A,Hossain:2014xha}; for plateau and hilltop potentials, which align inflationary predictions with the measured spectral index \cite{2025EPJC...85...48M,Dimopoulos:2017zvq}; and for non-minimally coupled scalars, where curvature coupling modifies both inflationary dynamics and late-time evolution \cite{10.1093/ptep/ptaf128}. A related research direction seeks similar unification within modified gravity rather than through a dedicated field. For example, in $F(R)$ gravity, a single geometric sector can accommodate inflation as well as early and late dark energy epochs \cite{PhysRevD.103.044036}, and such models are continually tested against the latest observational data \cite{ODINTSOV2026100471}. However, this scenario faces a significant challenge. Because the potential lacks a minimum, the inflaton cannot reheat the Universe through oscillations, necessitating alternative mechanisms such as inefficient gravitational particle production or instantaneous preheating via coupling to a secondary field \cite{Felder:1999pv}. This difficulty does not arise in the construction presented below, where radiation is generated by the decay of the waterfall condensate rather than by the inflaton, and the inflaton continues to dissipate into the resulting thermal bath throughout the second stage of inflation.

This paper explores three lines of theoretical development. We analyze a scalar sector coupled to the Gauss–Bonnet invariant \(\mathcal{G}\), where the coupling function is determined by the vacuum expectation values of the relevant fields rather than a fixed constant. The potential lacks a minimum, ensuring that the inflaton remains dynamical from the inflationary epoch to the present. Initially, the Gauss–Bonnet term dominates the dynamics, producing a super-inflationary phase with $\dot H>0$, which we solve in closed form. The resulting blue-tilted spectrum naturally explains the observed deficit of power at low multipoles. This phase concludes through a geometric mechanism where the Gauss–Bonnet invariant decreases, the stabilization of a transverse waterfall field is lost, leading to symmetry breaking that both ends accelerated expansion and, via condensate decay, generates the Universe's first radiation. This approach resolves a longstanding challenge in quintessential inflation, as the inflaton does not require reheating through oscillations about a non-existent minimum. The generated radiation establishes initial conditions for a subsequent, prolonged warm phase. We derive the scalar and tensor spectra for this stage, retaining higher-curvature contributions, and demonstrate that inflaton fluctuation growth is governed by the bath's shear viscosity. This framework makes a quartic potential viable and constrains the parameter space through internal consistency rather than the measured spectral index, with which it generally agrees. After inflation, the field evolves along the exponential tail to serve as the dark energy component. Additionally, a third scalar, released near matter–radiation equality, transforms a small mass scale into a large Gauss–Bonnet coupling with a galactic-scale associated length, exerting negligible influence on background expansion. Consequently, the dark sector arises from the same geometric structure that drives inflation, obviating the need for an independent relic population or a separate cosmological constant.

The paper is organized as follows. Section~\ref{sec2} introduces the theoretical framework, detailing the scalar potential, the Gauss-Bonnet coupling function, and the notation adopted throughout. Section~\ref{sec3} addresses the super-inflationary phase, including its background evolution, phantom-like kinematics, the resulting perturbation spectra, and the geometric mechanism responsible for its termination. Section~\ref{sec4} treats the transition between the two inflationary stages and the warm phase that follows. It begins with the waterfall transition, the development of the transverse instability and the coherent oscillations of the waterfall condensate whose decay produces the first radiation, and then turns to the warm inflationary phase itself, where the microphysical origin of the dissipation coefficient, the background dynamics, the derivation of the scalar and tensor spectra with the Gauss-Bonnet contributions retained, and the role of shear viscosity in controlling the growth of inflaton fluctuations. The numerical results and the viable parameter-space region are presented at the end of the section. Section~\ref{sec5} explores the post-inflationary era, including the release of the third scalar, the emergence of the large coupling, and its interpretation as a length scale. Section~\ref{sec:conclusion} summarizes the main findings, discusses their implications, and outlines open questions.

\raggedbottom
\section{The inflationary model}
\label{sec2}
Let us consider the following action as a first step in our study
\begin{equation}
     S=\int_{}^{} d^4x\sqrt{-g}\left[\frac{{M_p^{2}}}{{2}}R+f(\phi,\chi,\sigma)\mathcal{G}-\frac{1}{2}g^{\mu\nu}K_{\mu\nu}-V(\phi,\chi,\sigma)\right],
    \label{eq1}
\end{equation}
where \(R\) is the Ricci scalar, \(\mathcal{G}=R^{2}-4R_{\mu \nu } R^{\mu \nu }+R_{\mu \nu \rho \sigma } R^{\mu\nu \rho \sigma }\) is the Gauss-Bonnet invariant. \(K_{\mu\nu}=\sum\partial_{\mu}\phi_i\partial_{\nu}\phi_i\) is the kinetic energy of the three fields where \(\phi_i\in\{\phi,\chi,\sigma\}\) and the scalar field \(\phi\) is what we consider as the quintessential inflaton, \(V(\phi,\chi,\sigma)\) is the bare scalar potential and \(f(\phi,\chi,\sigma)\) is a coupling function to the invariant \(\mathcal{G}\). The action (\ref{eq1}) is a subclass of Horndeski’s theory for the coupling to one scalar field, providing the general scalar–tensor theory description \cite{Kobayashi_2019}. However, for multi-scalar fields, we can refer to \cite{2025arXiv251115423K} for this kind of study. 

By varying the action (\ref{eq1}) with respect to the metric tensor \(g_{\mu\nu}\) with the flat FLRW ansatz, and assuming that all the scalar fields are homogeneous, we obtain the following Friedmann equations
\begin{equation}
    3M_{p}^{2}H^{2}=\frac{1}{2}\sum\dot{\phi}^2_i+V(\phi,\chi,\sigma)-24\dot{f}(\phi,\chi,\sigma)H^3,
   \label{eq2}
\end{equation}
\begin{equation}
    3M_{p}^{2}H^{2}+2M_{p}^{2}\dot{H}+\frac{1}{2}\sum\dot{\phi}^2_i-V(\phi,\chi,\sigma)+16H\left(H^2+\dot{H}\right)\dot{f}(\phi,\chi,\sigma)+8H^2\ddot{f}(\phi,\chi,\sigma)=0,
   \label{eq3}
\end{equation}
where \(H=\dot{a}/a\) is the Hubble parameter, the Ricci scalar along with the Gauss-Bonnet invariant expressed as \(R=12H^2+6\dot{H}\) and \(\mathcal{G}=24H^2(H^2+\dot{H})\), respectively.

Within the framework of our investigation of quintessential inflation, aimed at unifying the early time inflationary epoch with the late time dark energy, we consider a bare potential \(V\propto \exp{(-Q\phi)}\) and coupling function \(f\propto \exp{(Q\phi)}\). A primary challenge that arises in this framework is the reconciliation of the vastly different energy scales required at early and late times, a high-energy scale capable of driving inflation and an extremely low-energy scale consistent with present day dark energy, together with the need for a viable transition to a radiation dominated epoch compatible with Big Bang Nucleosynthesis (BBN).

Moreover, existing studies of models based on Gauss–Bonnet gravity with both exponential coupling and exponential potential indicate that such configurations generally fail to provide a smooth and natural exit from the inflationary phase, as noted in \cite{PhysRevD.88.123508}. A similar difficulty appears in the strong-gravity regime when the Gauss–Bonnet term is present with a coupling function but in the absence of a scalar potential. In this case, the dynamics do not allow a graceful transition to our observational universe; instead, the system either remains trapped in an eternally inflating phase or evolves toward a Milne-like phase \cite{PhysRevD.92.041302}.

In addition, the Gauss–Bonnet term raises a fundamental question about its role in modifying gravitational dynamics: whether its effects are confined to high-energy regimes or can also induce nontrivial contributions at cosmological scales.

In the context of four-dimensional Gauss–Bonnet gravity \cite{PhysRevLett.124.081301}, studies of cosmological perturbations indicate that the theory propagates a scalar degree of freedom in addition to the tensor modes, while vector perturbations remain absent. This feature is consistent with current observational constraints. Similarly, in models based on Gauss–Bonnet gravity applied to the formation of cosmic structures \cite{HAGHANI2020100720,WANG2021100813}, it has been shown that modifications of gravity can effectively lead to a time dependent correction to the effective Newton’s constant, thereby enhancing the growth of structures.

However, a major issue with Gauss–Bonnet gravity, whether with a scalar field coupling or a constant coupling parameter, is the difficulty of fixing the coupling magnitude and identifying the relevant energy scale of the interaction. As emphasized in \cite{Fernandes_2022}, different studies adopt widely varying values depending on the specific physical regime under consideration, which may suggest that the coupling is effectively scale-dependent.

On the other hand, since the Gauss–Bonnet invariant decreases rapidly during the cosmological evolution of the Universe (\(\mathcal{G}\sim H^4\)), maintaining a significant modification of gravity at late times typically requires a sufficiently large coupling function to compensate for this suppression. However, such large couplings are strongly constrained by Solar System tests of gravity. It is therefore important to distinguish between local and cosmological scales. In high-curvature environments such as the Solar System, the Gauss–Bonnet term can be significant even with relatively small couplings, whereas at cosmological scales a larger effective coupling may be required. This tension suggests that a viable framework may require a screening mechanism to reconcile local gravity constraints with cosmological effects.

Referring back to the review \cite{Fernandes_2022}, if one assumes a constant Gauss--Bonnet coupling $\tilde{\alpha}$, maintaining a dynamically significant contribution at late times requires a large magnitude of $\tilde{\alpha} \sim 10^{52} \, \mathrm{m}^2$. This corresponds to a characteristic length scale of $\sqrt{\tilde{\alpha}} \approx 3.1 \, \mathrm{Gpc}$, which is of the order of the cosmological horizon. However, a survey of observational bounds across different cosmological epochs reveals that the allowed magnitude of the coupling must be strictly smaller at early times. This tension strongly indicates that any phenomenologically viable coupling should not be constant, but must instead grow dynamically as the Universe evolves.

Such dynamical behavior is naturally realized in models featuring a field-dependent coupling, such as $f(\phi) = f_0 e^{Q\phi}$ \cite{Chahboun_2024,Chahboun_2026}. In this framework, the present-day coupling parameter can be associated with galactic scales rather than the cosmological horizon. For instance, a coupling value of $f_0 \sim 4.2 \times 10^{26} \, M_\odot \, \mathrm{kpc \, (km/s)}^2$ yields a length scale of $\sqrt{\tilde{f}_0} \approx 3.3 \, \mathrm{kpc}$ (where $\tilde{f}_0 = f_0 M_p^2 / 2$). At this scale, the dimensionless background contribution evaluates to $f_0 H_0^2 / M_p^2 \sim 10^{-13}$. While subdominant at the background level, this interaction remains perturbatively significant, providing a viable mechanism to slightly enhance the growth of cosmic structures.

To resolve the issues of the graceful exit from inflation and the problematic magnitude of the Gauss-Bonnet coupling, we propose a hybrid inflation scenario \cite{PhysRevD.49.748,PhysRevD.49.6410,RevModPhys.78.537,BASTEROGIL2009157}. In this framework, we adopt a bare potential and a coupling function motivated by Grand Unified Theories (GUTs) scale, which are defined as follows
\begin{subequations}
\begin{eqnarray}
 &V(\phi, \chi, \sigma)=\left[ V_0 + \frac{\lambda_\chi}{4}\left(\chi^2 - \nu_\chi^2\right)^2 + \gamma^4\chi^4\phi^4\left(1-\frac{\sigma^4}{\nu_\sigma^4}\right)+ \frac{\lambda_\sigma}{4}\left(\sigma^2 - \nu_\sigma^2\right)^2 \right] e^{-Q\phi}, \\
 &f(\phi, \chi, \sigma)=\left[ f_1\frac{g^4}{4} \left(\chi^2 + \nu_\chi^2\right)\left(\chi^2 - 4\nu_\chi^2\right) + 2f_2g^4\chi^4 \left( 1 - \frac{\sigma^2}{\Lambda^2} + \frac{\sigma^4}{M_s^4} \right) \right] e^{nQ\phi},
\end{eqnarray}
\end{subequations}
where \(V_0\) is the late-time dark energy scale, the scalar field \(\chi\) represents the waterfall field responsible for the waterfall mechanism. We assumes that the \(\chi\) sector is a GUT energy scale \(\lambda_\chi=10^{-2}\),\(\nu_\chi=10^{-2}M_p\) and \(g\) is a normalization constant related to an inverse mass scale also in the GUT energy scale (\(g\equiv1/M_g\)), \(f_1\) and \(f_2\) are dimensionless constants (in the SI system they are of dimension of \([M][L]^3[T]^{-2}\)) of order unity \(\mathcal{O}(1)\) with \(f_1=f_2\). The scalar field \(\sigma\) is a light scalar field related to the matter-dominated epoch and \(\Lambda\sim M_p\). The \(\gamma^4\)-potential will play a crucial role in driving inflation, and we will discuss it in more detail later.

\section{Super inflation phase}
\label{sec3}
In the previous section, we established the theoretical framework of our inflationary model, specifying explicit forms for both the scalar potential and the Gauss-Bonnet coupling function. In particular, we introduced a Mexican-hat type potential for the fields $\chi$ and $\sigma$, designed to realize a hybrid inflation scenario and trigger spontaneous symmetry breaking. This potential is further coupled to the scalar field $\phi$, which also acts as a quintessence-like field at late times. Additionally, we proposed a functional form for the coupling that depends on the fields $\chi$ and $\sigma$, allowing its magnitude to evolve dynamically throughout cosmic evolution. In this way, the coupling naturally adapts to the model's different dynamical regimes.

Building on this, we now investigate the model's dynamical viability. To this end, we focus on the scalar sector involving the inflaton $\phi$ and the waterfall field $\chi$, while ignoring the \(\sigma\) contribution compared to the other fields during all the inflationary phase. By varying the action (\ref{eq1}) with respect to $\phi$ and $\chi$, we obtain the corresponding Klein-Gordon equations

For the $\phi$ scalar field
\begin{equation}
    \ddot{\phi} + 3H\dot{\phi} + \partial_\phi V_{\mathrm{eff}} = 0,
\label{eq5}
\end{equation}

For the $\chi$ scalar field
\begin{equation}
    \ddot{\chi} + 3H\dot{\chi} + \partial_\chi V_{\mathrm{eff}} = 0,
\end{equation}
where $\partial_{\phi_i}$ denotes the partial derivative with respect to the corresponding scalar field, and the effective potential is defined as $V_{\mathrm{eff}} = V - \mathcal{G}f$.

During the initial inflationary phase, we assume that the quintessential inflaton $\phi$ begins at a large field value. In this regime, the waterfall field $\chi$ acquires an effective mass $m_{\chi}$ that is extremely large compared to the Hubble parameter ($m_{\chi} \gg H$). Consequently, the field $\chi$ is rapidly trapped stably at the minimum of its effective potential (i.e., $\chi \simeq 0$). Evaluated at this origin, the effective squared mass of the waterfall field takes the following form
\begin{equation}
    m_{\chi\simeq0}^2 = \frac{3}{2}f_1g^4\nu_\chi^2 e^{nQ\phi}\mathcal{G} - \lambda_\chi\nu_\chi^2 e^{-Q\phi}.
    \label{eq7}
\end{equation}

Regarding the coupling function \(f(\phi,\chi,\sigma)\), we introduce a real parameter \(n>1\) into the exponential expression $\exp(nQ\phi)$ rather than assuming the simplest case of \(n=1\). This distinction is critical because, while the waterfall mechanism is employed to guarantee the exit from inflation, the specific choice of \(n=1\) permits a successful phase transition only in the case of pure de Sitter phase where the Hubble parameter \(H\) is constant (and consequently, a constant \(\mathcal{G}\)). However, a realistic quintessential inflation scenario is inherently dynamical; as the inflaton rolls, the Universe undergoes a quasi-de Sitter evolution. Under these realistic physical conditions, in order for the effective mass of the waterfall field to eventually become tachyonic and trigger the necessary symmetry breaking.

Returning to the Friedmann equations, we assume that the quintessential inflaton $\phi$ is slowly rolling down its effective potential. In this regime, the standard slow-roll approximations hold: \(|\dot{H}|\ll H^2\), \(\dot{\phi}^2 \ll V\), and \(|\ddot{\phi}| \ll 3H|\dot{\phi}|\). Furthermore, regarding the Gauss-Bonnet contribution, we impose the constraint \(8H|\dot{f}|/M_p^2 \ll 1\). This last condition ensures that while the Gauss-Bonnet coupling dynamically stabilizes the waterfall field, it does not completely dominate the background expansion dynamics of the Universe. 

Under these conditions along with \(\chi=0\), the equations (\ref{eq2}) and (\ref{eq3}) becomes
\begin{equation}
    3M_p^2H^2\simeq \frac{M^4}{4}e^{-Q\phi},
    \label{eq8}
\end{equation}
\begin{equation}
    2M_p^2\dot{H}\simeq-\dot{\phi}^2-8nQf_{in}\dot{\phi}H^3e^{nQ\phi},
    \label{eq9}
\end{equation}
where \(M^4=\lambda\nu_\chi^4\) and \(f_{in}=f_1g^4\nu_\chi^4\). Consequently, the bare potential along with the coupling function, are given by \(V=M^4e^{-Q\phi}/4\) and \(f=-f_{in}e^{nQ\phi}\), respectively. 

Substituting the slow-roll conditions into the inflaton equation (\ref{eq5}), together with the approximation \(\mathcal{G} \simeq 24H^4\), we obtain
\begin{equation}
    \dot{\phi}\simeq Q\left(\frac{M^4}{12H}e^{-Q\phi}-8nf_{in}H^3e^{nQ\phi}\right).
    \label{eq10}
\end{equation}

Referring back to the inflationary paradigm and its connection to observations, measurements of the Cosmic Microwave Background (CMB) indicate a suppression of power at large angular scales, corresponding to low multipoles $\ell\lesssim30$ \cite{Planck2018VII,Schwarz:2015cma}. In standard slow-roll inflation, characterized by $\dot{H}<0$, the primordial power spectrum is predicted to be nearly scale-invariant with a slight red tilt ($n_s<1$), which does not naturally account for this observed deficit of power on the largest scales.

One possible mechanism proposed to address this issue involves the so-called super-inflation phase, characterized by $\dot{H}>0$ \cite{Piao:2004uq,Biswas_2014}. In such a regime, the dynamics of horizon crossing are modified, and the resulting spectrum of perturbations can exhibit a blue tilt ($n_s>1$). If this phase occurs only for a limited number of e-folds before the standard slow-roll era, it can affect only the largest observable scales, suppressing power at low multipoles while preserving the successful predictions of conventional inflation at smaller scales.

Within the framework of standard cosmology, realizing a super-inflationary scenario ($\dot{H}>0$) is notoriously difficult, as it inherently requires the introduction of exotic matter components that violate the null energy condition (NEC). To circumvent this pathology and achieve a growing Hubble parameter, the theoretical framework typically necessitates ultraviolet (UV) corrections to the dynamics, such as the quantum gravity effects seen in Loop Quantum Cosmology (LQC) \cite{Ashtekar:2011ni,2005LRR.....8...11B} or non-singular bouncing cosmologies \cite{Brandenberger:2016vhg,Novello:2008ra}. The primary cosmological advantage of such a UV-driven super-inflationary phase is its ability to generate a strongly blue-tilted spectrum at large scales, naturally suppressing the anomalous low multipoles in the CMB before standard, red-tilted slow-roll inflation takes over in the infrared (IR) \cite{Biswas_2014,PhysRevD.87.063518}. In the present model, this mechanism is realized naturally via the Gauss--Bonnet coupling, a highly motivated higher-curvature UV correction to General Relativity.

Specifically, the waterfall field \(\chi\) must remain stably trapped at the origin. Imposing the stability condition into the equation (\ref{eq7}) \(m_{\chi\simeq0}^2 \geq 0\) together with the equation (\ref{eq10}), we get to the following condition
\begin{equation}
    \dot{\phi} \leq Q(3-8n)f_{in}H^3e^{nQ\phi}.
\end{equation}

Based on the previous requirement \(n>1\), the factor \((3-8n)\) is strictly negative. As a result, the field velocity is strictly negative (\(\dot{\phi} < 0\)). Consequently, the quintessential inflaton \(\phi\) decreases toward smaller field values, causing the background potential energy to grow over time. This continuous growth forces the Hubble parameter \(H\) to increase (\(\dot{H} > 0\)), naturally driving the Universe into a phase of super-inflation.

To advance our analysis, we refer back to Eq. (\ref{eq7}). As the inflaton \(\phi\) decreases, the dynamics remain stable until the effective squared mass of the waterfall field \(m_{\chi\simeq0}^2\) vanishes at the critical point \(\phi_c\), given by 
\begin{equation}
    \phi_c = \frac{1}{(n-1)Q} \ln \left( \frac{4M_p^4}{f_{in}M^4} \right).
    \label{eq12}
\end{equation}

Having established the analytical expression for the critical point \(\phi_c\), we now turn our attention to the coupling parameter \(g\). To constrain its value, an upper bound naturally arises from the requirement that the critical field value \(\phi_c\) be positive. Ensuring that the super-inflationary phase ends before the inflaton crosses into the negative region (\(\phi_c \geq 0\)) yields a maximum limit
\begin{equation}
    g \leq \frac{\sqrt{2}M_p}{f_1^{1/4}\lambda_\chi^{1/4}\nu_\chi^2},
\end{equation}
this evaluates to an upper bound of \(g_{max} \approx 4.47\times10^{4} M_p^{-1}\), where we assumed \(f_1 = 1\).

Conversely, to establish the lower bound, we rely on the dynamics of the inflaton field \(\phi\). Since the field is strictly decreasing, we impose the condition \(\phi_c < \phi_*\), where \(\phi_*\) is the value of the field when the large observable scales cross the horizon. Using the expression for \(\phi_c\) from Eq. (\ref{eq12}), we obtain the lower bound on \(g\) as follows
\begin{equation}
    g>\frac{\sqrt{2}M_p}{f_1^{1/4}\lambda_\chi^{1/4}\nu_\chi^2}e^{\frac{(1-n)Q\phi_*}{4}}.
\end{equation}

While the lower bound depends on both \(Q\) and the parameter \(n\), that means the corresponding lower bound on the coupling \(f_{\mathrm{in}}\) varies with the choice of \(n\). Concerning the upper bound, if one takes \(g = g_{max}\), this implies the relation \(4 M_p^4 = f_{in} M^4\), which in turn requires that inflation ends at the critical point \(\phi_c = 0\). However, under our assumption that the scalar field remains positive throughout the inflationary phase, this scenario becomes inconsistent. In particular, since the super-inflationary phase is expected to last only for a short time \(\sim 7\) e-folds, its exit at \(\phi_c = 0\) would prevent the realization of a subsequent standard inflationary phase lasting at least \(50\) e-folds while maintaining a monotonically decreasing field.

For this reason, we adopt a benchmark value of \(g\) smaller than \(g_{max}\), while still satisfying the lower bound, and fix \(g = 4.4 \times 10^{4}\, M_p^{-1}\) throughout the inflationary epoch. Given that \(n > 1\), we further impose a physicality condition on the evolution of the scalar field, \(\phi_* \leq 10 M_p\). Combining this requirement with the chosen benchmark value of \(g\), we derive an additional constraint on \(n\) that ensures consistency with both the bounds on \(g\) and the field range. Numerically, we find \(n_{min} \approx 2.33\) based on the chosen value of \(g\) and \(Q=0.005M_p^{-1}\).

The duration of super-inflationary phase is defined dynamically as follows
\begin{equation}
    N=\int_{\phi_*}^{\phi_c}\frac{H}{\dot{\phi}}d\phi.
    \label{eq14}
\end{equation}

By substituting the expressions for $H$ from Eq. (\ref{eq8}) and $\dot{\phi}$ from Eq. (\ref{eq10}), the analytical form for $N$ is given by 
\begin{equation}
N=-\frac{\phi_*}{QM_p^2}+\frac{1}{(n-1)Q^2M_p^2}\ln{\left(\frac{8ne^{(n-1)Q\phi_*}-\frac{12M_p^4}{f_{in}M^4}}{8n-3}\right)}.
\label{eq16}
\end{equation}

\begin{figure}[H]
\centering
\begin{tabular}{c c}
    \includegraphics[width=0.52\textwidth]{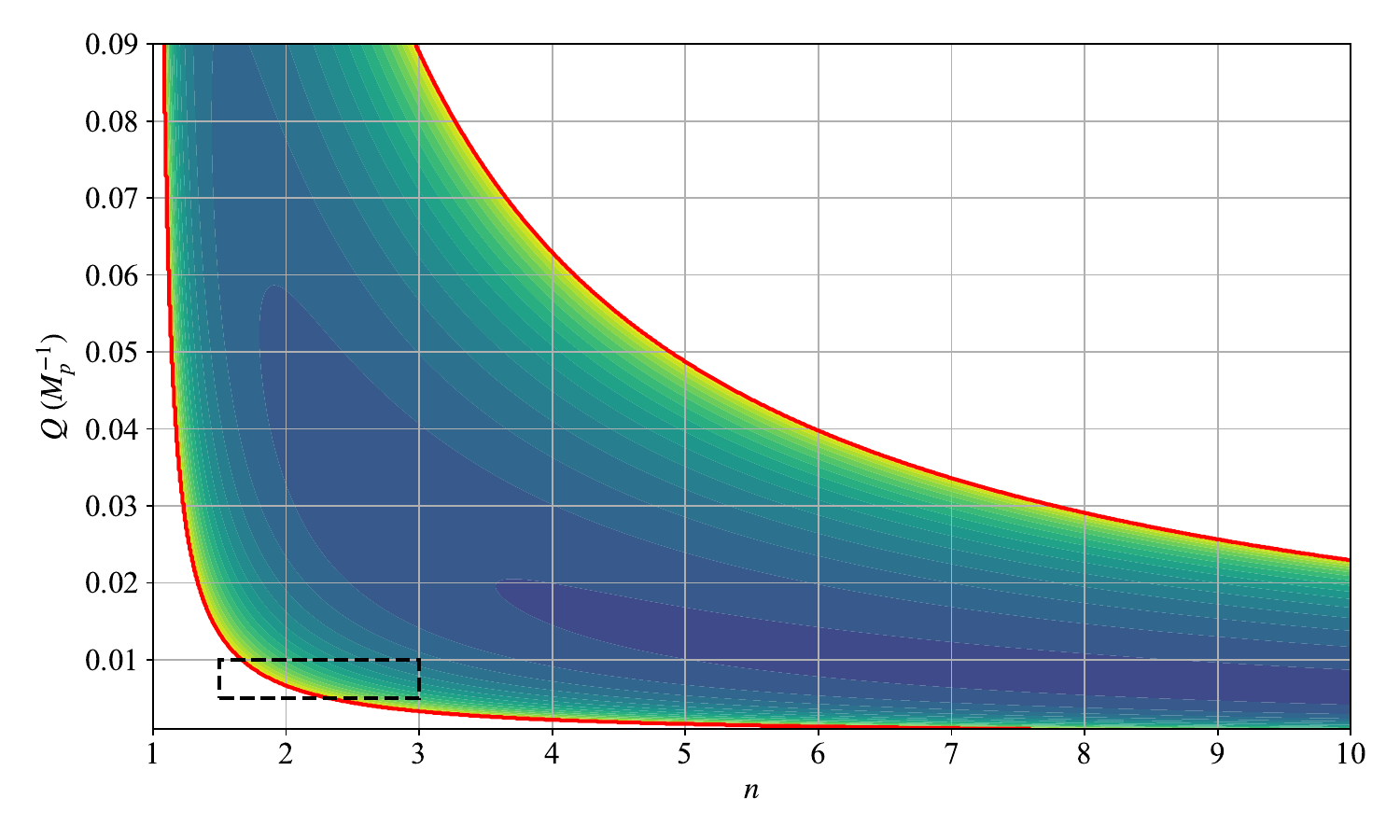} &
    \includegraphics[width=0.52\textwidth]{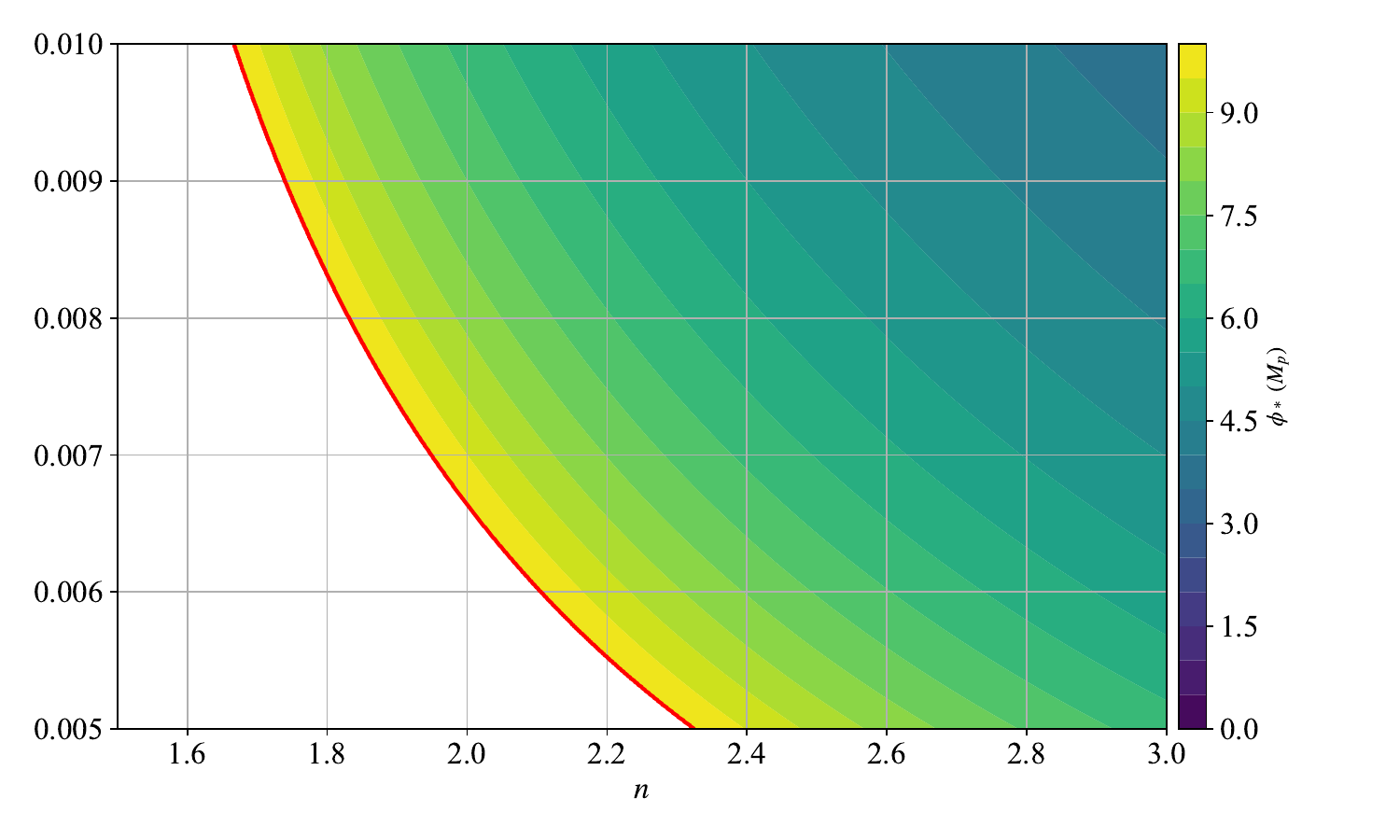}
\end{tabular}
\caption{The viability space for \((n, Q)\) plane. The colored contour indicates the assumption of the initial field value \(\phi_*\) to achieve \(7\) e-folds of super-inflation. The region bounded by the red line represents the space where the physical constraints \(\phi_* \leq 10 M_p\) and \(\phi_c\geq0\) are satisfied.}
\label{fig1}
\end{figure}

In the Figure \ref{fig1}., we present the viability space of \((n,Q)\) based on our assumptions of \(n_{min}\approx2.33\), \(g = 4.4 \times 10^{4}\, M_p^{-1}\) and \(\phi_*\leq10M_p\), yielding $f_{in}M^4\simeq3.74M_p$. To ensure the waterfall mechanism triggers below $\phi_*$, we find the lower bound $(n-1)Q > 0.0065M_p^{-1}$.

\begin{figure}[H]
\centering
\begin{tabular}{c c}
    \includegraphics[width=0.51\textwidth]{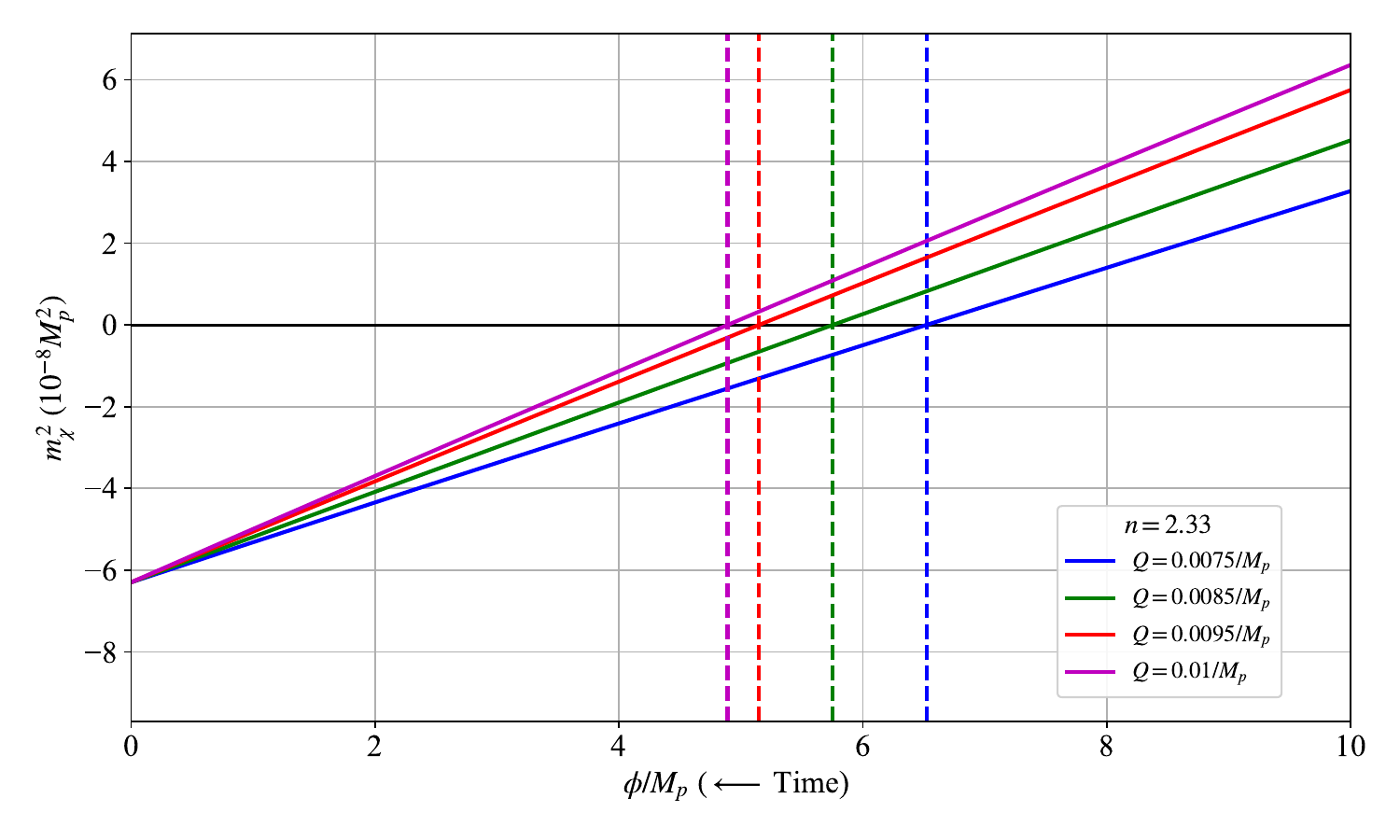}&
    \includegraphics[width=0.51\textwidth]{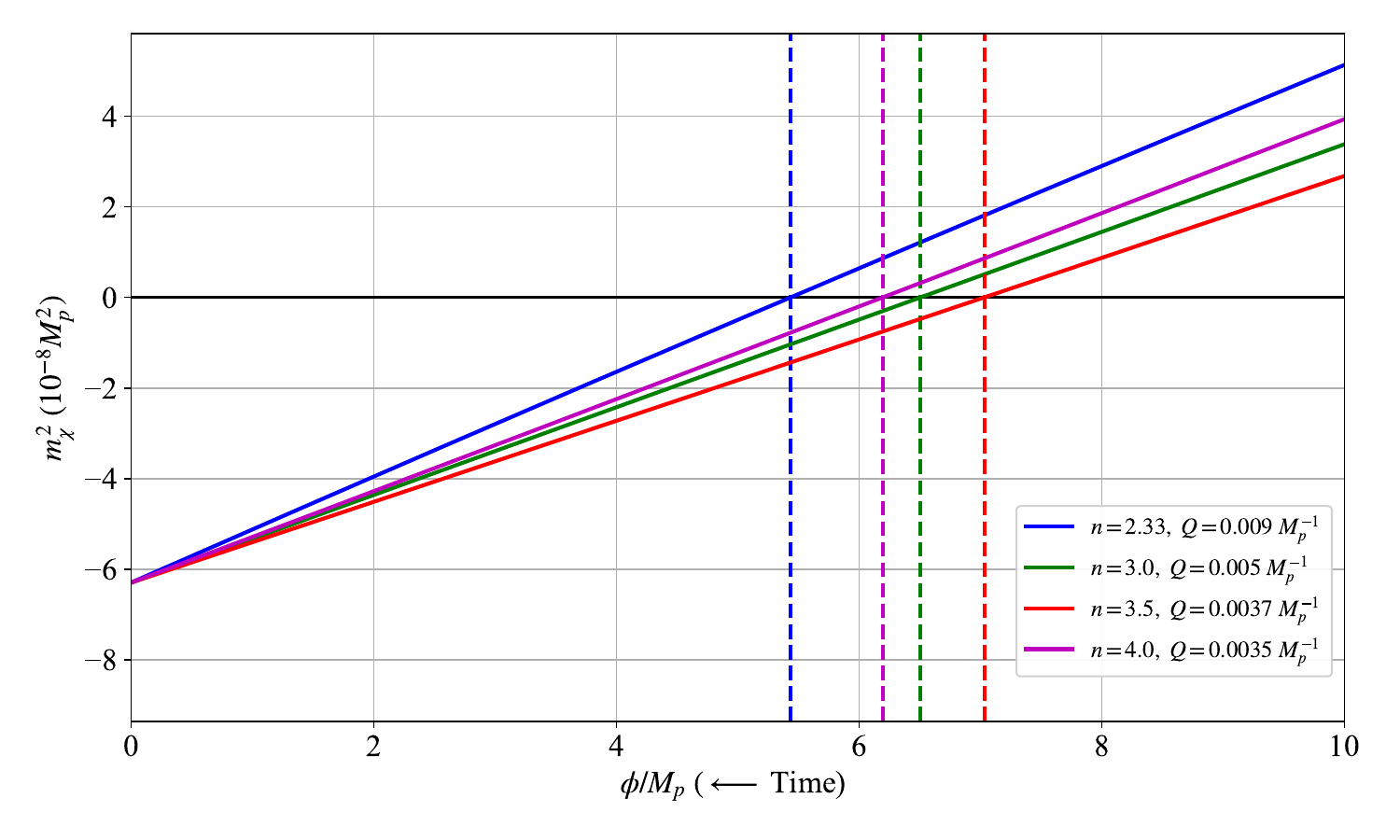}\\
\end{tabular}
\caption{The effective mass $m_\chi^2$ at the origin, Eq.~(\ref{eq7}). Left: fixed $n=2.33$ for several values of $Q$. Right: a comparison across viable combinations of $(n,Q)$. Dashed vertical lines mark $\phi_{c}$, where the effective mass vanishes and the waterfall is triggered.}
    \label{fig2}
    \end{figure}

The evolution of the effective mass of the waterfall field \(m^2_{\chi\simeq0}\) for several choices of the parameters \(n\) and \(Q\) is presented in Figure \ref{fig2}.
The effective mass traces the competition between the two terms of Eq.~(\ref{eq7}): the Gauss--Bonnet contribution scales as $e^{(n-2)Q\phi}$ and the tachyonic one as $e^{-Q\phi}$, so their ratio falls as $e^{(n-1)Q\phi}$ as the field rolls. Stabilisation is therefore lost at $\phi_{c}$, a single unambiguous crossing that triggers the waterfall. The four cases shown have different $n$ and $Q$ but comparable $(n-1)Q$, and their critical points are ordered accordingly rather than by $n$ alone.

Let us now shift our focus to one of the central pillars of the inflation studies which is the primordial power spectrum. This observable constitutes a cornerstone of the present analysis, particularly in the context of the super-inflationary phase and its potential role in addressing the low-multipole anomalies. Accordingly, we proceed by investigating the primordial spectrum within the standard perturbative framework, considering both scalar and tensor modes.

Firstly, a leading order, we analyze metric perturbations in the scalar sector, adopting the scalar gauge decomposition of a flat FLRW as follows
\begin{equation}
    ds^2=-(1+2\varphi)dt^2+a^2(t)(1-2\psi)\delta_{ij}dx^i dx^j,
    \label{eq17}
\end{equation}
where \(\varphi\) and \(\psi\) are the scalar potentials.

In this choice of gauge, the gauge-invariant comoving curvature perturbation \(\mathcal{R}\) is given by
\begin{equation}
    \mathcal{R}=\psi+\frac{H}{\dot{\phi}}\delta\phi.
    \label{eq18}
\end{equation}

To track the quantum fluctuations and their evolution, it is highly convenient to transition to Fourier space, where \(\nabla^2\to-k^2\), and work with the conformal time \(d\tau=dt/a\). Upon the canonical field redefinition \(v\equiv z_s\mathcal{R}\), the coupled system of perturbed Einstein and Klein-Gordon equations collapses to a single second-order equation of Mukhanov-Sasaki type \cite{1985ZhPmR..41..402M,10.1143/PTP.76.1036} (we refer to \cite{1998PhLB..437..284K,1999PhLB..460...41K,PhysRevD.71.063536,PhysRevD.96.124022} for a complete treatment of the perturbed field equations in the presence of a Gauss-Bonnet coupling),
\begin{equation}
    v'' + \left( c_s^2 k^2 - \frac{z_s''}{z_s} \right) v = 0,
    \label{eq19}
\end{equation}
where a prime denotes differentiation with respect to \(\tau\), equation \eqref{eq19} describes a harmonic oscillator with a time-dependent frequency. Its two limiting regimes carry a transparent physical meaning: deep inside the sound horizon (\(c_sk\gg aH\)) the gradient term dominates, the modes oscillate adiabatically, and the Bunch-Davies vacuum \(v_k\to e^{-ic_sk\tau}/\sqrt{2c_sk}\) may be imposed as an initial condition \cite{10.1098/rspa.1978.0060}; after sound-horizon crossing the effective mass \(z_s''/z_s\) takes over, the growing solution dominates, and \(\mathcal{R}\) freezes on super-horizon scales.

Two quantities thus carry the entire influence of the higher-curvature coupling on the scalar sector; the propagation speed \(c_s^2\) and the field \(z_s\). The corresponding expressions have been obtained in \cite{PhysRevD.88.123508,PhysRevD.98.083521} for an arbitrary coupling function, and we specialise them to the present model.

Along with \(\epsilon_1=-\dot{H}/H^2\) and \(\epsilon_{2}=\frac{d\ln{|\epsilon_1|}}{d\ln{a}}\), it is convenient to introduce the dimensionless Gauss-Bonnet flow parameter
\begin{equation}
    \delta_1 = -\frac{8\dot{f}H}{M_p^2},
    \qquad
    \delta_2 = \frac{d\ln|\delta_1|}{d\ln a} = \frac{\dot{\delta_1}}{H\delta_1},
    \label{eq20}
\end{equation}
which measures the rate of change of the coupling in Hubble units and vanishes identically in the minimally coupled limit. Assuming $\ddot{f}\ll H\dot{f}$, the sound speed in terms of $\delta_1$ reads \cite{Guo:2010jr,PhysRevD.88.123508,PhysRevD.71.063536,PhysRevD.98.083521}
\begin{equation}
    c_s^2=\frac{\dfrac{2\dot{\phi}^2}{M_p^2}(1-\delta_1)^2
    +\delta_1^2\left[3H^2+4\dot{H}-4\delta_1(H^2+\dot{H})\right]}
    {\dfrac{2\dot{\phi}^2}{M_p^2}(1-\delta_1)^2+3\delta_1^2H^2(1-\delta_1)},
    \label{eq21}
\end{equation}
while the field \(z_s\), whose square is built on the same combination is given by
\begin{equation}
    z_s^2=\frac{a^2M_p^2}{H^2(2-3\delta_1)^2}
    \left[\frac{4\dot{\phi}^2}{M_p^2}(1-\delta_1)^2+6\delta_1^2H^2(1-\delta_1)\right].
    \label{eq22}
\end{equation}

Both expressions reduce to their canonical single-field counterparts, \(c_s^2\to1\) and \(z_s^2\to a^2\dot{\phi}^2/H^2\), when \(\delta_1\to0\), as they must. We note that \(c_s^2>0\) and \(z_s^2>0\) throughout the super-inflationary phase, so that the scalar sector is free of both ghost and gradient instabilities.

The effective mass term controls the scale dependence of the resulting spectrum. Using the quasi-de Sitter relation \(aH\simeq-[(1-\epsilon_1)\tau]^{-1}\), valid to first order in the flow parameters, Eq.~\eqref{eq22} yields
\begin{equation}
    \frac{z_s''}{z_s}=\frac{1}{\tau^2}\left(\nu^2-\frac{1}{4}\right),
    \qquad
    \nu=\frac32+\epsilon_1+\frac{2\epsilon_1\epsilon_2-\delta_1\delta_2}
    {2\left(2\epsilon_1-\delta_1\right)}.
    \label{eq23}
\end{equation}

The brevity of the super-inflationary stage, which lasts only about seven e-folds, together with the near-constancy of the Gauss-Bonnet parameter over that interval, guarantees that the flow parameters (and hence \(\nu\)) may be treated as constant to excellent accuracy. Equation \eqref{eq19} then reduces to a Bessel equation, whose solution matching the Bunch-Davies vacuum in the remote past is \(v_k(\tau)=\tfrac{1}{2}\sqrt{\pi|\tau|}\,H^{(1)}_{\nu}(c_sk|\tau|)\), with \(H^{(1)}_{\nu}\) the Hankel function of the first kind. Evaluating \(\mathcal{P}_{\mathcal{R}}=(k^3/2\pi^2)|v_k/z_s|^2\) on super-horizon scales \(c_sk\ll aH\), where the asymptotic behavior of the Hankel function applies, the curvature power spectrum at horizon crossing \(c_sk= aH\) reads \cite{Guo:2010jr}
\begin{equation}
    \mathcal{P}_{\mathcal{R}}\simeq\frac{H^2}{4\pi^2M_p^2c_s^3}\left(2\epsilon_1-\delta_1\left(1+\epsilon_1-\delta_2\right)+\frac{3\delta_1^{2}}{2\left(1-\delta_1\right)}\right)^{-1},
    \label{eq24}
\end{equation}
whose scale dependence is characterised by the scalar spectral index defined below.

To further develop our perturbative analysis, and following the treatment of scalar perturbations, we now turn to the second physically relevant sector: tensor perturbations. These modes correspond to transverse and traceless fluctuations of the metric and are interpreted as gravitational waves. In the cosmological context, they are of particular importance as they encode primordial information about the early Universe, providing a direct probe of the inflationary epoch. The perturbed metric in the tensor gauge takes the following form
\begin{equation}
    ds^2=-dt^2+a^2(t)\left(\delta_{ij}+h_{ij}\right)dx^idx^j,
\end{equation}
where the transverse-traceless metric perturbation as \(\delta g_{ij}=a^2h_{ij}\) expanding in the two polarization states \(\lambda=+,\times\), the canonically normalized mode function \(u^{\lambda}_k=z_Th^{\lambda}_{k}/2\) obeys an equation of the same structure as Eq.~\eqref{eq19} \cite{PhysRevD.88.123508,PhysRevD.98.083521}
\begin{equation}
    \frac{d^2u^{\lambda}_k}{d\tau^2}
    +\left(c_T^2k^2-\frac{z_T''}{z_T}\right)u^{\lambda}_k=0,
    \label{eq26}
\end{equation}
In terms of the Gauss-Bonnet framework, the gravitational waves velocity \(c_T\) and the field \(z_T\) are given by \cite{PhysRevD.71.063536,PhysRevD.96.124022} 
\begin{equation}
    z_T^2=a^2M_p^2\left(1-\delta_1\right),
    \label{eq27}
\end{equation}
\begin{equation}
    c_T^2=1 + \frac{8(\ddot{f} - H\dot{f})}{M_p^2 + 8H\dot{f}}
\end{equation}
Both expressions reduce to their Einstein-gravity counterparts, \(z_T^2\to a^2M_p^2\) and \(c_T^2\to1\), in the limit \(f\to0\) (\(\delta\to0\)), and the absence of tensor ghosts and gradient instabilities requires \(c_T^2>0\), a condition that is satisfied throughout the super-inflationary phase. It is worth stressing that, unlike the scalar sector where the deviation of the sound speed from unity is of second order, \(c_s^2-1=\mathcal{O}(f^2)\), the tensor modes acquire a correction already at first order in the coupling, \(c_T^2-1=\mathcal{O}(f)\).

Proceeding as in the scalar case, the effective mass entering Eq.~\eqref{eq26} takes the form as follows
\begin{equation}
    \frac{z_T''}{z_T}=\frac{1}{\tau^2}\left(\mu^2-\frac14\right),
    \qquad
    \mu=\frac32+\epsilon_1.
    \label{eq29}
\end{equation}

Imposing the Bunch-Davies vacuum \cite{10.1098/rspa.1978.0060} and evaluating the mode function in the super-horizon limit $c_Tk\ll aH$, where the asymptotic form of the Hankel function applies, the tensor power spectrum, evaluated at horizon crossing $c_Tk=aH$, reads \cite{Guo:2010jr}
\begin{equation}
    \mathcal{P}_T\simeq\frac{2H^2}{\pi^2M_p^2\left(1-\delta_1\right)c_T^3},
    \label{eq30}
\end{equation}
whose scale dependence is characterised by the tensor spectral index
\begin{equation}
    n_T=\frac{d\ln\mathcal{P}_T}{d\ln k}=3-2\mu=-2\epsilon_1.
    \label{eq31}
\end{equation}

Since the horizon-crossing conditions \(c_sk=aH\) and \(c_Tk=aH\) differ only at second order in the flow parameters, this distinction may be consistently neglected \cite{Guo:2010jr}. Combining Eqs~\eqref{eq30} and the scalar amplitude then yields the tensor-to-scalar ratio
\begin{equation}
    r=\frac{\mathcal{P}_T}{\mathcal{P}_{\mathcal{R}}}
    \simeq16\epsilon_1-8\delta_1,
    \label{eq32}
\end{equation}
while the scalar spectral index follows from Eq.~\eqref{eq24} as
\begin{equation}
    n_s-1=\frac{d\ln\mathcal{P}_{\mathcal{R}}}{d\ln k}=3-2\nu
    =-2\epsilon_1-\frac{2\epsilon_1\epsilon_2-\delta_1\delta_2}{2\epsilon_1-\delta_1}.
    \label{eq33}
\end{equation}
Equations \eqref{eq31}, \eqref{eq32} and \eqref{eq33} make explicit the fact that the Gauss-Bonnet coupling supplies an additional degree of freedom, \(\delta_1\), which enters \(r\) but not \(n_T\). The standard single-field consistency relation \(r=-8n_T\) is therefore violated, and is replaced in the present model by a relation involving the coupling strength.

\begin{figure}[H]
\centering
\begin{tabular}{c c}
    \includegraphics[width=0.51\textwidth]{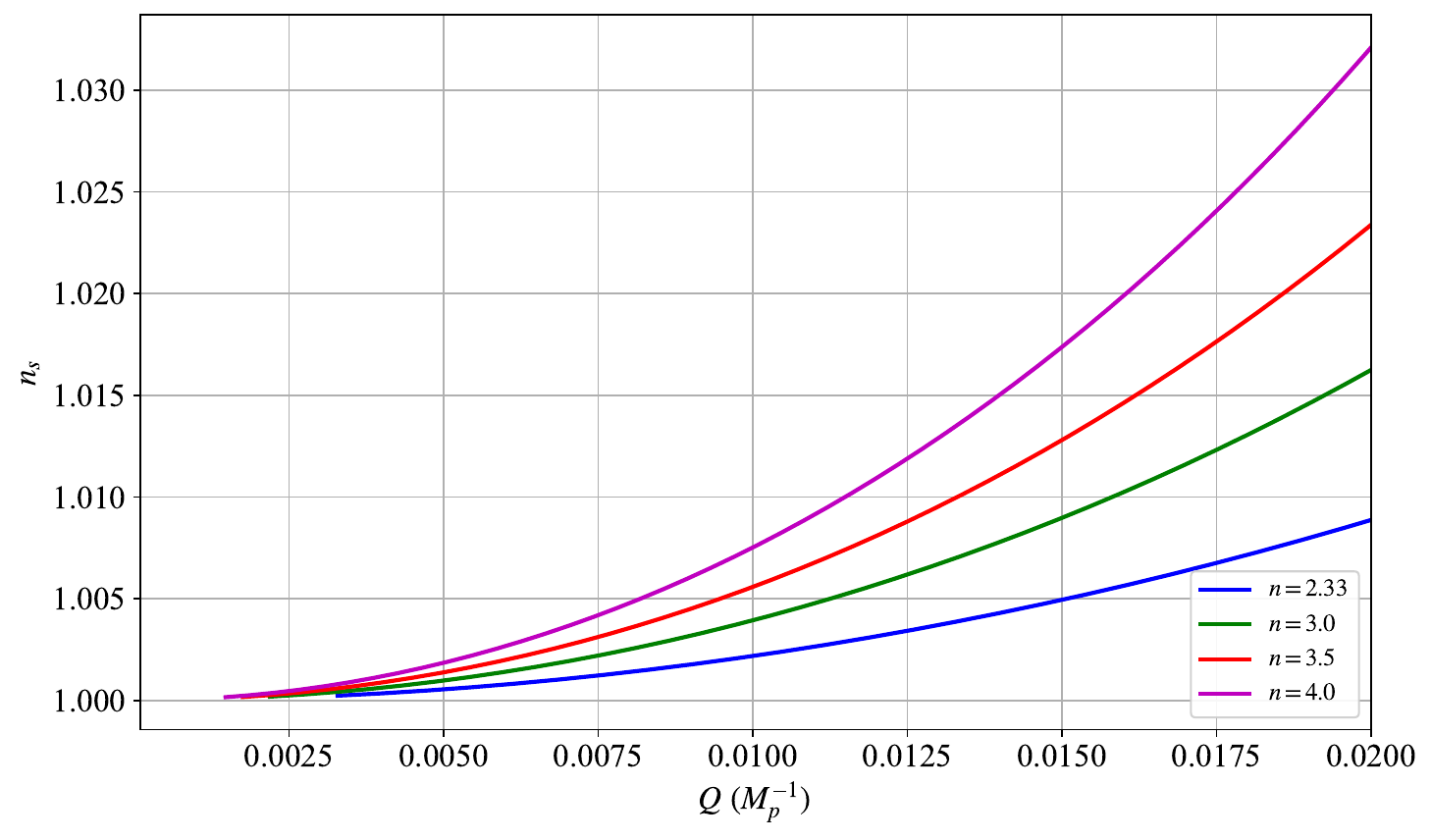}&
    \includegraphics[width=0.51\textwidth]{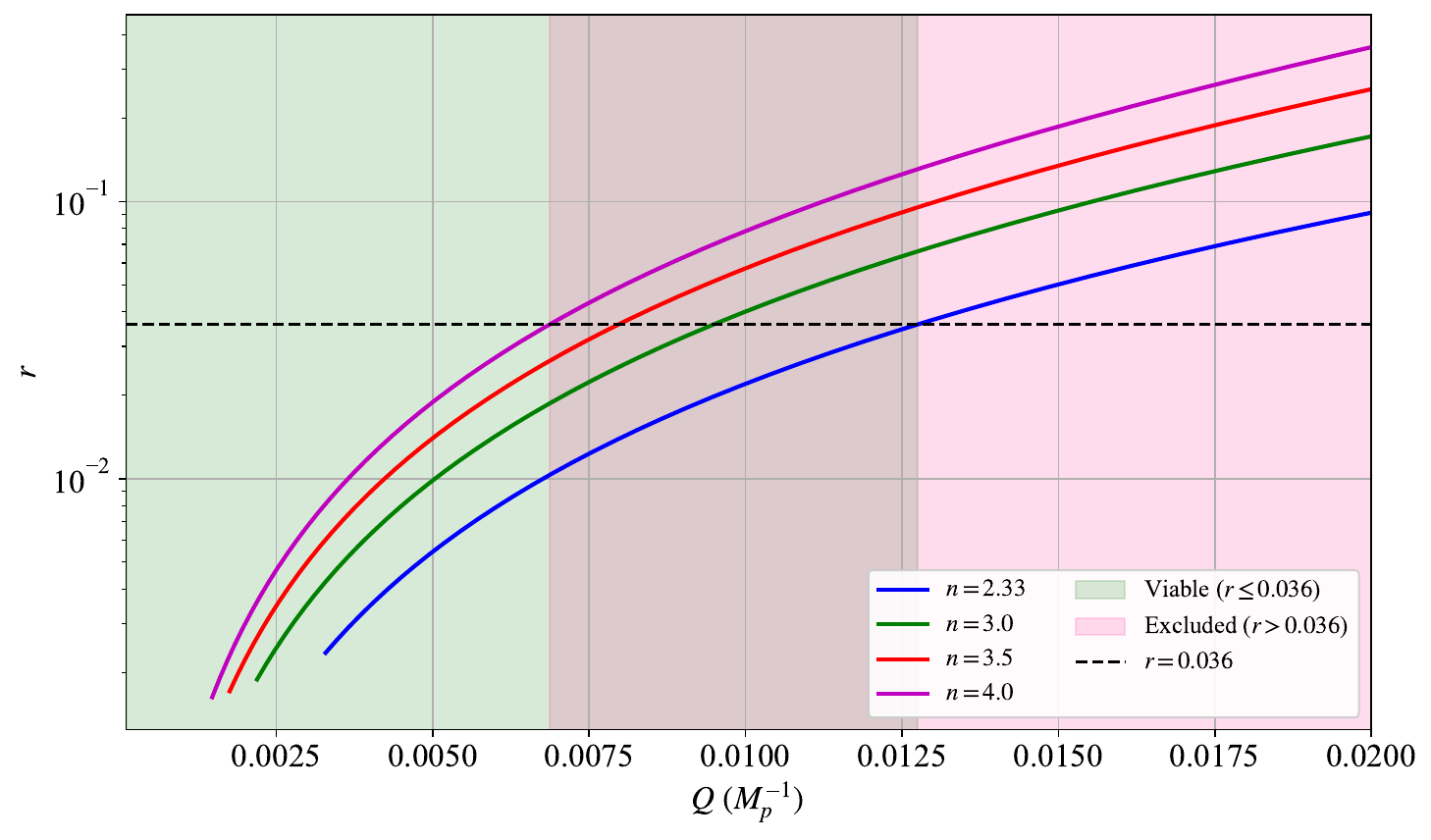}\\
\end{tabular}
\caption{Behavior of the inflationary observables as functions of the potential slope \(Q\), for four values of the parameter \(n\), evaluated at horizon crossing \(N_*=7\) e-folds before the waterfall transition. Left panel: scalar spectral index \(n_s\) from Eq.~(\ref{eq34}). Right panel: tensor-to-scalar ratio \(r\) from Eq.~(\ref{eq35}), shown on a logarithmic scale. The dashed horizontal line marks the current upper bound \(r<0.036\) \cite{PhysRevLett.127.151301}. The green band indicates the region in which all the models displayed satisfy this bound, the pink band the region in which all of them violate it, and the overlap the intermediate range where viability depends on \(n\).}
    \label{fig3}
    \end{figure}

Combining the general expressions for $n_s$ and $r$ in Eqs.~(\ref{eq32}) and (\ref{eq33}) with the expression of \(N\) given in Eq.~(\ref{eq16}), and defining $\Delta\phi = \phi_* - \phi_c$, we obtain the following explicit expressions at horizon crossing for \(N_* = 7\) $e$-folds:

\begin{equation}
    n_s=1+Q^2M_p^2\left[2(n-1)+\frac{(2n-1)(8n-3)}{3}\exp\left((n-1)Q(N_*QM_p^2+\Delta\phi)\right)\right],
    \label{eq34}
\end{equation}
\begin{equation}
    r = 8Q^2M_p^2\left[\frac{8n-3}{3}\exp\left((n-1)Q(N_*QM_p^2+\Delta\phi)\right)\right]^{2}.
    \label{eq35}
\end{equation}
These analytical expressions allow us to compute the scalar spectral index $n_s$ and the tensor-to-scalar ratio $r$ over \(Q\), whose behavior at horizon crossing is shown in Figure \ref{fig3}.

Figure \ref{fig3}. displays the predictions of Eqs.~\eqref{eq34} and \eqref{eq35} across the parameter space. Because the super-inflationary phase lasts only 7 e-folds, the exponential factor common to both observables departs from unity by less than one percent, so that \(n_s-1\) and \(r\) scale to a very good approximation as \(Q^2\), with a mild steepening at the upper end of the range. The monotonic ordering with \(n\) has an equally simple origin, a larger exponent corresponding to a stronger Gauss-Bonnet domination at horizon crossing and hence to a larger \(|\epsilon_1|\). The spectrum is blue throughout, a feature inherited from the hybrid character of the model, but its tilt remains small, \(n_s-1\lesssim4\times10^{-3}\) over the viable region, so that the spectrum stays very nearly scale invariant. This is essential for the consistency of the two-phase scenario, since the modes leaving the horizon during this brief burst correspond to the largest observable scales. The two panels are moreover not independent: the slope \(Q\) cancels in the ratio \(r/(n_s-1)\), which varies by less than two percent across the whole range of exponents considered. Consequently, although the four curves cross \(r=0.036\) at values of \(Q\) spanning nearly a factor of two, from \(Q\simeq6.9\times10^{-3}M_p^{-1}\) for \(n=4\) to \(Q\simeq1.3\times10^{-2}M_p^{-1}\) for \(n=2.33\), the spectral index takes essentially the same value \(n_s\simeq1.0035\) at each crossing. The observational bound on the tensor amplitude therefore translates, within this scenario and largely independently of the model parameters, into the constraint \(n_s-1\lesssim3.5\times10^{-3}\).

To close the super-inflationary phase, we must specify the mechanism that terminates it, since this stage does not end spontaneously. When $\phi$ evolves independently, the hierarchy between the two forces in the effective potential remains constant, and $\epsilon_1<0$ persists indefinitely. Termination is achieved through the waterfall field, whose transverse effective mass \eqref{eq7} vanishes at $\phi=\phi_c$. As $\phi$ approaches $\phi_c$, the mass $m_\chi^2$ approaches zero, and $\chi$ transitions from a heavy spectator fixed at the origin to a light degree of freedom primarily influenced by Hubble friction. For $\phi<\phi_c$, $\chi$ becomes tachyonic, resulting in rapidly growing accumulated fluctuations. Linearization near the critical point yields $\eta_\chi\equiv-m_\chi^2/H^2\simeq12(M_p/\nu_\chi)^2(n-1)Q\,(\phi_c-\phi)$, and with $\nu_\chi\sim10^{-2}M_p$, the prefactor is approximately $10^4$. Consequently, $\eta_\chi$ reaches unity after a displacement of order $10^{-3}M_p$, within a few times $10^{-2}$ e-folds. The waterfall is thus classified as fast, in contrast to scenarios characterized by extended trajectories along which observable modes exit during the waterfall itself \cite{PhysRevD.83.063518}. The modes relevant to the spectra computed above exit the horizon well before the transition, ensuring that the waterfall does not influence the predictions for $n_s$, $r$, and $n_T$. During this process, $\phi$ remains effectively frozen, as its velocity displaces the field by only a few percent of $\phi_c$, thereby maintaining a quasi-de Sitter background. As a result, $\chi$ evolves as a spectator within a fixed geometry.

What terminates the phase is not the growth of $\chi$ but that of its kinetic energy. Eliminating the potential between \eqref{eq2} and \eqref{eq3}, the first Hubble flow parameter reads
\begin{equation}
    \epsilon_1=\frac{\dot\phi^{2}+\dot\chi^{2}+8H^{2}\ddot f-8H^{3}\dot f}
    {2H^{2}\left(M_{p}^{2}+8H\dot f\right)},
    \label{eq36}
\end{equation}
in which the higher-curvature terms are negative and drive super-inflation, whereas the kinetic contributions are strictly positive. Since $\phi$ is frozen at $\phi_c$, the coupling varies only through the waterfall field, so that $\dot f=f_\chi\dot\chi$ and $\ddot f=f_{\chi\chi}\dot\chi^2+f_\chi\ddot\chi$, and near the origin $f_\chi\simeq-\tfrac32(f_{\rm in}/\nu_\chi^2)\,\chi\,e^{nQ\phi_c}$, so that $f_{\chi\chi}=f_\chi/\chi$. Every term in $\epsilon_1$ then carries the same factor $\chi^2$, the amplitude of the growing mode cancels, and the vanishing of $\epsilon_1$ becomes a condition on the growth rate alone. It is met when the waterfall fluctuations grow at half the Hubble rate, that is at $\eta_\chi=7/4$, a value independent of the coupling strength and of the symmetry-breaking scale. The sequence of events is as follows. Geometric stabilisation weakens as $\phi$ descends, $\chi$ becomes light and subsequently tachyonic at $\phi_c$, and the kinetic energy of the growing fluctuations satisfies the aforementioned condition after a further displacement of order $10^{-3}M_p$, causing the expansion to transition from super-inflationary to ordinary within a fraction of an e-fold and transporting $\chi$ to the vicinity of its true vacuum in slightly more than one e-fold.
\section{Warm inflation phase}
\label{sec4}
In the previous section, we investigated a short phase of super-inflation within the framework of hybrid inflation. In this setup, the scalar field $\chi$ provides a suitable mechanism for exiting this phase through a tachyonic instability. Building on this picture, we derived constraints on the model parameters governing this transient regime and analyzed the associated inflationary observables, in particular the scalar spectral index $n_s$ and the tensor-to-scalar ratio $r$. We show that this phase can account for the behavior at low multipoles, with a spectrum characterized by a mild blue tilt while remaining close to scale invariance.

As the scalar field $\chi$ approaches its true vacuum, the initial stage of inflationary dynamics concludes. This event does not end inflation entirely. Rather, it marks the end of the first, brief phase within a broader inflationary evolution. The following, extended phase of warm inflation becomes dominant. Before examining its realization in the present model, it is instructive to review the general framework.

The warm inflation scenario was introduced in \cite{PhysRevLett.74.1912}, developed in \cite{PhysRevLett.75.3218} as an alternative to the conventional scenario, where the inflaton is assumed to be completely decoupled during accelerated expansion. The central premise is that, since the inflaton must ultimately transfer its energy to the Standard Model, there is no fundamental reason to prohibit this transfer during inflation. Allowing such a transfer modifies the equation of motion by introducing a dissipative friction term $\Gamma_\phi\dot\phi$, and energy conservation then requires that the same coefficient sources a radiation bath, $\dot\rho_r+4H\rho_r=\Gamma_\phi\dot\phi^2$. If dissipation exceeds dilution, a quasi-thermal bath is maintained throughout inflation, preventing the Universe from becoming cold. The strength of this effect is quantified by $Q_{\rm wi}=\Gamma_\phi/3H$, with the weak and strong dissipative regimes corresponding to $Q_{\rm wi}\ll1$ and $Q_{\rm wi}\gg1$, respectively.

Three key consequences arise, each pertinent to the present discussion. First, the friction term becomes $3H(1+Q_{\rm wi})$ instead of $3H$, thereby relaxing the slow-roll conditions and allowing potentials that are too steep for vacuum inflation to become viable \cite{Berera:2008ar}. Second, thermal rather than vacuum fluctuations generate perturbations, with their amplitude set by the fluctuation-dissipation theorem rather than the uncertainty principle. This change affects both the normalization and the shape of the spectrum \cite{BERERA2000666,Hall:2003zp}; specifically, the scalar amplitude is enhanced while the tensor amplitude remains unchanged, suppressing the tensor-to-scalar ratio by powers of $Q_{\rm wi}$. Third, because inflation concludes with radiation already present, the Universe transitions smoothly into the radiation era without requiring a separate reheating stage \cite{Berera:2008ar}.

The main challenge in realizing this scenario is that a coupling strong enough to sustain the radiation bath typically induces thermal corrections of order $cT^2$ to the inflaton mass. When $T>H$, these corrections compromise the flatness of the potential. The now-standard solution is to render the interaction indirect: the inflaton couples to heavy mediators that are never thermally populated, and only these mediators interact with the light fields \cite{Berera:2008ar,Berera:2007qm,BasteroGil:2012cm}. In the warm little inflaton framework \cite{BasteroGil:2016qru}, a discrete interchange symmetry protects the inflaton mass to all orders. Integrating out the mediators in the low-temperature regime yields a dissipation coefficient that is cubic in temperature \cite{BasteroGil:2011mr}. The present analysis adopts this class of model, with the radiation bath generated not by an assumed initial condition but by the decay of the previously described waterfall condensate.

Returning to the \(\chi\)-sector, the waterfall field $\chi$ settles near its true vacuum $\chi_{\min}$ and undergoes coherent oscillations around this minimum. After overshooting the minimum, $\chi$ does not remain stationary but oscillates with an effective mass $m_\chi$ determined by the curvature of the potential at the minimum, typically with $m_\chi \gg H$. Consequently, damped harmonic motion describes its dynamics. The oscillating condensate of $\chi$ acts as a decaying matter component, transferring part of its energy to light degrees of freedom through its couplings. This decay produces the initial radiation, establishing a thermal bath during a brief sub-oscillatory phase.

First of all, we need to analyze the full potential of the \(\chi\)-sector given by the following expression
\begin{equation}
    U=\left[\frac{\lambda_\chi}{4}\left(\chi^2 - \nu_\chi^2\right)^2 + \gamma^4\chi^4\phi^4\left(1-\frac{\sigma^4}{\nu_\sigma^4}\right)\right] e^{-Q\phi}.
    \label{eq37}
\end{equation}

However, the scalar field $\sigma$ acquires a large mass from the Gauss-Bonnet invariant, $m_{\sigma,\mathcal{G}}^{2}\propto\mathcal{G}/\Lambda^{2}$, which remains large throughout inflation and holds $\sigma$ at its false vacuum $\sigma\simeq0$. The corresponding factor in $U$ is therefore unity to excellent accuracy, and the transverse sector reduces to a competition between the symmetry-breaking hat and the $\gamma^{4}-$portal. The bracket in $U$ is a parabola for \(\chi^2\), thus draws the true vacuum to
\begin{equation}
    \chi_{\rm min}=\frac{\pm\nu_{\chi}}{\sqrt{1+\mathcal{R}_\chi(\phi)}},    
    \label{eq38}
\end{equation}
where \(\mathcal{R}_\chi(\phi)=4\gamma^{4}\phi^{4}/\lambda_{\chi}\). 

The displacement depends on $\phi$ is, at first sight, an unwelcome feature: where $\mathcal{R}_\chi$ of order unity, the vacuum expectation value of $\chi$ would drift as the inflaton rolls, and every quantity built from it, including the Gauss-Bonnet coupling function, which we wish to depend on $\phi$ solely through the exponential $e^{nQ\phi}$, would inherit a spurious field dependence. We now show that this does not occur.

The requirement follows from demanding that the transition be a genuine waterfall, that is, that it liberate a substantial fraction of the vacuum energy. Since the portal vanishes identically at the origin, the potential difference between the false and true vacuum is given by
\begin{equation}
    \Delta U=U(\chi=0)-U(\chi_{\rm min})
    =\frac{M^4}{4(1+\mathcal{R}_\chi)}e^{-Q\phi}.
    \label{eq39}
\end{equation}

For $\mathcal{R}_\chi\gg1$, the true vacuum rises to meet the false one, the symmetry breaking becomes energetically trivial, and no appreciable radiation is produced. The mechanism's consistency therefore requires $\mathcal{R}_\chi\lesssim1$. This is a necessary condition to have  $\chi_{\rm min}\simeq\nu_\chi$.

The decisive statement comes instead from the magnitude of $\gamma$ and from the fact that $\mathcal{R}_\chi$ depends on it through the fourth power. The coupling $\gamma$ is not a free parameter. However, \(\gamma\) is fixed by the observed amplitude of the curvature perturbation, which the warm phase must produce and then $\gamma\leq10^{-2}M_p^{-1}$, this is verifiable by taking the optimal case where the filed \(\chi\) sit exactly at the true vacuum of the hat potential \(\nu_\chi\), therefore the potential becomes \(\propto\gamma^4\nu_\chi^4\phi^4\), and to ensure the continuity of the Hubble rate between the two phases we need \(H_{\rm wi}(\phi_c)\leq H_{\rm si}(\phi_c)\) (where \(H_{\rm si}\) and \(H_{\rm wi}\) are respectively the rates in the super inflation phase and the warm phase) and with $\lambda_{\chi}=10^{-2}M_p$ we find that \(\gamma\leq3.33(n-1)Q\), and with the fact that \(Q\) is less than \(10^{-2}M_p^{-1}\) we find \(\gamma(n,Q)\leq10^{-2}M_p^{-1}\) with \((n,Q)\) dependence, combine all of this we get to the bound on \(\mathcal{R}_\chi\) to keep \(\chi_{\rm min}\simeq\nu_\chi\) as follows 
\begin{equation}
    \mathcal{R}_\chi(\phi)\leq4\times10^{-6}\left(\frac{\phi}{M_{p}}\right)^{4}.
    \label{eq40}
\end{equation}

Two consequences follow. First, $\chi_{\rm min}=\nu_{\chi}\bigl[1+\mathcal{O}(\mathcal{R}_\chi)\bigr]$, so the waterfall field settles at its symmetry-breaking scale to within a per cent and may be taken to sit there exactly. Second, nothing built from $\chi$, in particular the Gauss-Bonnet coupling, inherits any $\phi$ dependence beyond the explicit exponential, and the structure $f\propto e^{nQ\phi}$ used throughout is preserved. Since $\mathcal{R}_\chi$ is largest at the transition and $\phi_{c}$ depends on $n$ and $Q$ only through $(n-1)Q$, this holds in a definite region of parameter space, where we proceed with $\chi_{\rm min}=\nu_{\chi}$ and keeping $\mathcal{R_\chi}$ below the assumed bound. Before continuing to find the region of parameter space of \(n,Q\) where \(\chi_{\rm min}\simeq\nu_\chi\), we first need to analyze the oscillation of the waterfall field \(\chi\) around its minimum.

The fate of the waterfall field after symmetry breaking has been studied extensively in hybrid inflation, where the tachyonic instability drives an initially violent, non-perturbative stage of particle production called tachyonic preheating, before the condensate settles and decays \cite{RevModPhys.78.537,PhysRevD.79.063531,PhysRevLett.87.011601,PhysRevD.64.123517}. The efficiency of that stage, and the extent to which the subsequent evolution may be treated perturbatively, depends on the hierarchy between the waterfall mass and the expansion rate. In the present model, $m_{\chi}\gg H$ places us firmly in the regime of rapid, coherent oscillations about the true vacuum, and we accordingly describe the transition by the linearised dynamics of $\chi=\chi_{\rm min}+\delta\chi(t)$ coupled to the radiation it produces, where we treat the radiation as a phenomenological fluid, characterized by its energy density alone and by a temperature defined through $\rho_{r}=\tfrac{\pi^{2}}{30}g_{*}T^{4}$, rather than as a field in the action, where the the relativistic degrees of freedom at temperature \(T\) is assumed to be \(g_{*}(T)=100\). The waterfall field therefore obeys the following equations
\begin{equation}
     3M_{p}^{2}H^{2}=\rho_{r}+\frac{1}{2}\delta\dot\chi^{2}+\frac{1}{2}m_{\chi}^2\delta\chi^2-24f_{\chi}\delta\dot\chi H^{3},
     \label{eq41}
\end{equation}
\begin{equation}
    \delta\ddot\chi+\bigl(3H+\Gamma_{\chi}\bigr)\delta\dot\chi+m_\chi^2\delta\chi=0,
    \label{eq42}
\end{equation}
\begin{equation}
    \dot\rho_{r}+4H\rho_{r}=\Gamma_{\chi}\,\delta\dot\chi^{2},
    \label{eq43}
\end{equation}
in which the friction receives, besides the Hubble term, a contribution from the decay of the condensate into light degrees of freedom. Assuming a Yukawa coupling $y\,\chi\bar\psi\psi$, the decay width is defined as follows, together with the mass \(m_\chi^2\) at the true vacuum \(\chi_{\rm min}\)
\begin{equation}
    \Gamma_{\chi}=\frac{y^{2}}{8\pi}m_{\chi} ,
    \qquad
    m_{\chi}^{2}=2\lambda_{\chi}\nu_{\chi}^{2}e^{-Q\phi} ,
    \label{eq44}
\end{equation}
where we note that the curvature at the minimum is independent of $\mathcal{R}_\chi$: the extremum condition $(1+\mathcal{R}_\chi)\chi_{\rm min}^{2}=\nu_{\chi}^{2}$ removes it identically, so that the portal displaces the vacuum but leaves the mass there, and hence the decay rate, untouched. On the other hand, the full expression of the mass \(m_\chi^2\) also has a contribution from the Gauss-Bonnet invariant \(\mathcal{G}\) \(\propto f_{\chi\chi}\mathcal{G}\). However, this last term is negligible compared to the bare mass at the true vacuum after symmetry breaking, so \(m_\chi^2\) remains near the bare mass at the true vacuum. 

Furthermore, two dimensionless ratios characterize the resulting dynamics. The first, $m_{\chi}/H \sim 5 \times 10^{2}$, guarantees that the oscillations are coherent, with several hundred cycles occurring over about $2\text{--}3$ e-folds, so that the condensate behaves as a classical mode and tracks the minimum adiabatically as $\phi$ remains close to $\phi_c$. In this regime, the inflaton field is effectively frozen, with negligible kinetic energy, $\dot{\phi}^2 \ll V$, and thus its contribution to particle production through the coupling term $\Gamma_{\phi}\dot{\phi}^2$ can be safely neglected compared to the dominant decay of the $\chi$ condensate. The second ratio, $\Gamma_{\chi}/H \sim 10$, ensures that the waterfall field decays efficiently, transferring the energy (Eq.\eqref{eq47}) into relativistic species within less than a Hubble time, thereby initiating the formation of a radiation bath.
\begin{figure}[H]
\centering
\begin{tabular}{c c}
    \includegraphics[width=0.51\textwidth]{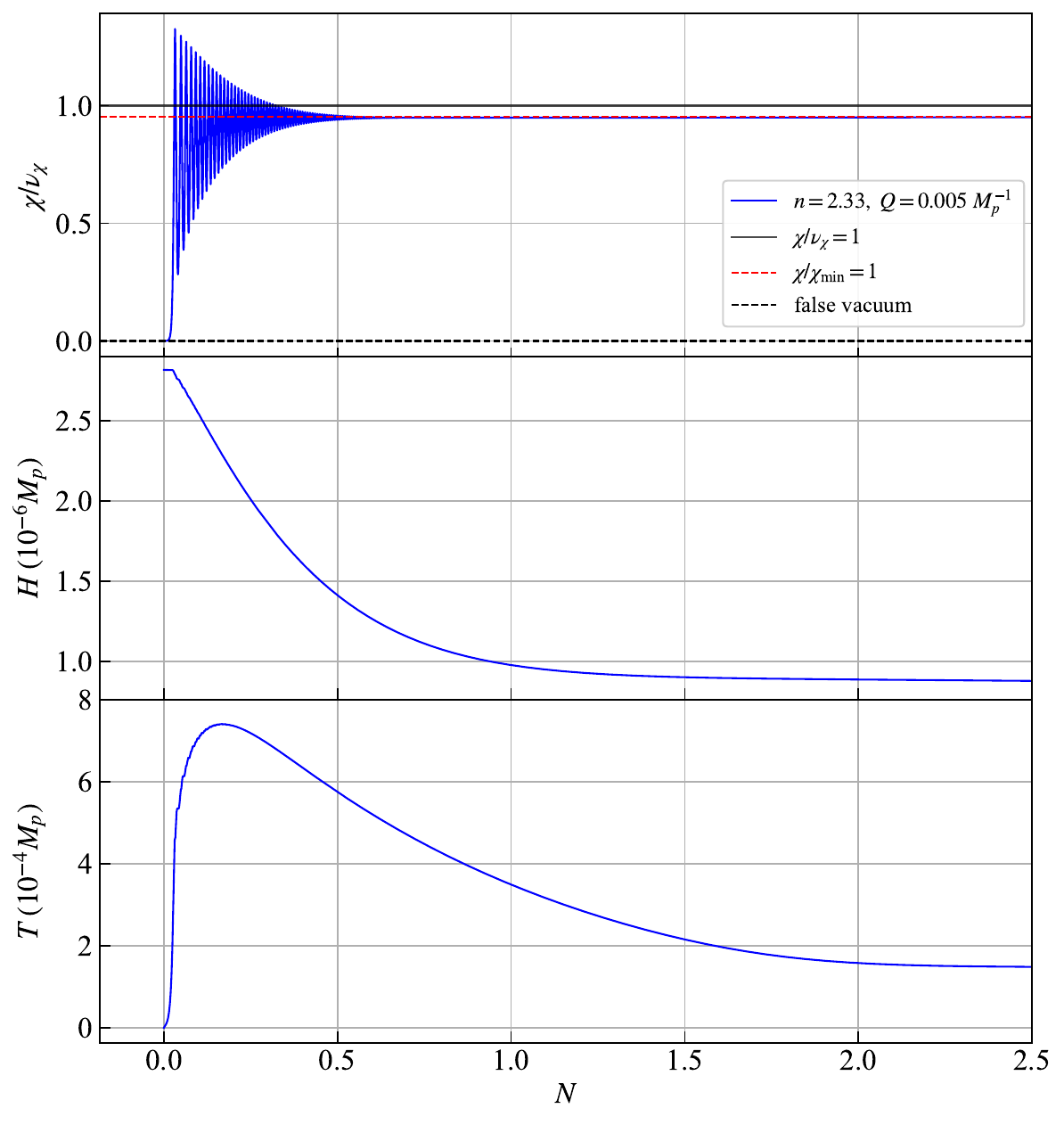}&
    \includegraphics[width=0.51\textwidth]{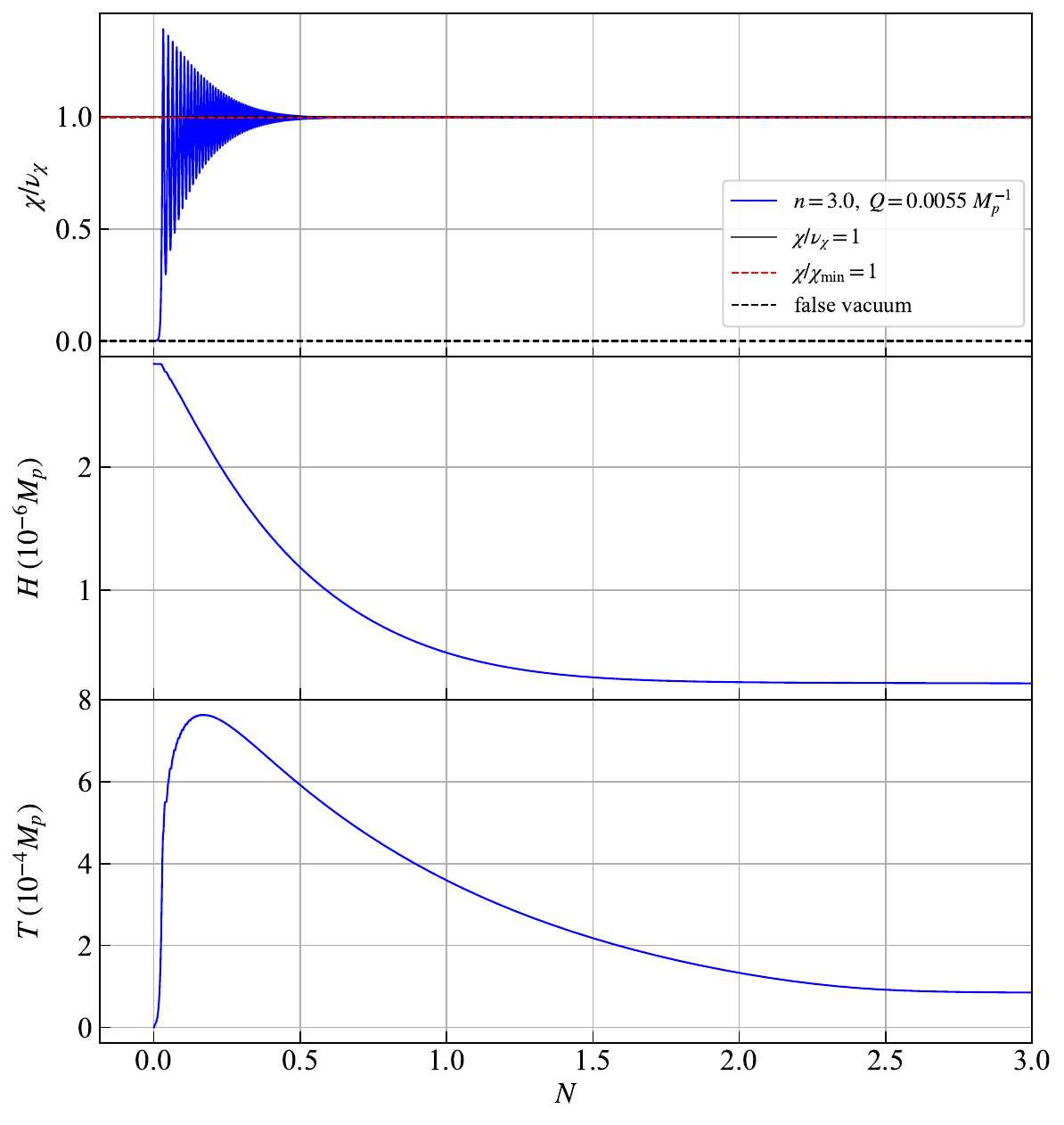}\\
\end{tabular}
\caption{Numerical evolution of the waterfall field, the Hubble rate and the radiation temperature during the oscillation of the waterfall field \(\delta\chi\), as functions of the number of e-folds $N$. Left column: $n=2.33$, $Q=0.005\,M_p^{-1}$, a combination for which the true minimum of the effective potential is displaced from the bare symmetry-breaking scale, $\chi_{\rm min}\neq\nu_\chi$. Right column: $n=3.0$, $Q=0.0055\,M_p^{-1}$, for which the two coincide. In the upper panels the solid grey line marks $\chi=\nu_\chi$, the dashed red line the actual minimum $\chi=\chi_{\rm min}$ reached by the field, and the dashed black line the false vacuum $\chi=0$.}
    \label{fig4}
    \end{figure}

In Fig.~\ref{fig4}, we illustrate the transition between the first and second phases by comparing two representative points in parameter space, $(n,Q) = (2.33,\,0.005\,M_p^{-1})$ and $(3,\,0.0055\,M_p^{-1})$, highlighting how the dynamics depends on the choice of $(n,Q)$. Each column's three panels show the same sequence of events. The waterfall field, released from the neighbourhood of the origin, traverses the full range to the true vacuum in less than \(1\) e-fold, then oscillates with an amplitude damped away within about half an e-fold; the oscillations are manifestly coherent and rapid on the expansion timescale, several tens of periods being completed per e-fold, which confirms $m_\chi\gg H$ and justifies treating the condensate as a non-relativistic matter component whose decay is perturbative. The Hubble rate falls by a factor of roughly three over the same interval, from its super-inflationary value to the plateau on which the warm phase settles, the drop occurring precisely while the oscillation amplitude is largest: it is the transfer of the vacuum energy liberated at the symmetry breaking, first into the kinetic energy of $\chi$ and then into radiation, that degrades the expansion rate. The temperature panels make the same point directly. Starting from a strictly vanishing value, the temperature rises steeply during the first few oscillations, peaks at $T\simeq7.5\times10^{-4}M_p$ around $N\simeq0.2$. It thereafter decreases smoothly as the source is exhausted and the expansion dilutes the bath. This peak is the signature of the condensate decay: the first quanta of the light sector are produced here, and the fact that $T$ is nonzero from this point onward, while identically zero before it, is what distinguishes the warm phase from the cold one that preceded it. We stress that at the peak one has $T/H\sim300$. Hence, the bath thermalizes on the expansion timescale by a wide margin, while $\rho_r/\rho_{\rm total}$ remains small which keeping the Universe stays potential-dominated throughout, and inflation continues, with the transition being a warming rather than a reheating. The subsequent slow decline of $T$ is not a return to a cold regime but an approach to the quasi-stationary balance between dissipative production and dilution that characterizes the warm inflationary attractor.

The two columns differ in one worth-noting respect. For the parameters in the right panel, the field settles exactly at $\chi_{\rm min}=\nu_\chi$, so the true vacuum coincides with the bare symmetry-breaking scale. For the parameters of the left panel, the minimum is displaced, $\chi_{\rm min}\simeq0.96\,\nu_\chi$, because the higher-curvature contribution to the effective potential does not vanish at $\chi=\nu_\chi$ and shifts the stationary point; a residual coupling therefore survives into the warm phase. The displacement is modest and the gross features of the transition, the duration of the oscillatory stage, the fall of $H$ and the shape of the temperature profile, are essentially unchanged between the two columns. Its significance is instead one of principle: only for combinations of $(n,Q)$ satisfying $\chi_{\rm min}=\nu_\chi$ does the scenario reduce, after the waterfall, to warm inflation.

Once the waterfall condensate has decayed completely, the Universe contains a bath of relativistic species while remaining potential-dominated, and inflation proceeds. It is necessary to analyze the behavior of each field in relation to this thermal bath, as a thermal mass correction $\Delta m^{2}\sim c_{\phi_i}T^{2}$ arises only from degrees of freedom that are light compared to the temperature and are thermally populated; contributions from heavier states are Boltzmann suppressed. The inflaton remains protected by the previously described construction, coupling exclusively to mediators with mass $m_{X}\gg T$, which prevents the induction of a correction $\sim c_\phi T^{2}$ and preserves the flatness of $V(\phi)$. In contrast, the waterfall field couples directly to the bath via its Yukawa interaction and consequently acquires $\Delta m_{\chi}^{2}\sim y^{2}T^{2}/6$. This correction remains manageable provided $c_{\chi}T^{2}<\lambda_{\chi}\nu_{\chi}^{2}$, or equivalently $T<\nu_{\chi}\sqrt{\lambda_{\chi}/c_{\chi}}$, ensuring that the symmetry remains broken and the transition is not reversed. For $\sigma$, the situation is reversed. During super-inflation, it is stabilized at the origin by the Gauss–Bonnet mass $m_{\sigma,\mathcal{G}}^{2}\propto\mathcal{G}/\Lambda^{2}$; however, this contribution decreases rapidly as $\mathcal{G}$ diminishes at the end of that phase, and the thermal bath subsequently provides the necessary stabilization through $c_{\sigma}T^{2}\sigma^{2}$, maintaining $\sigma\simeq0$ provided $T>\nu_{\sigma}\sqrt{\lambda_{\sigma}/c_{\sigma}}$. These competing requirements define a viable window $\nu_{\sigma}\sqrt{\lambda_{\sigma}/c_{\sigma}}\ll T<\nu_{\chi}\sqrt{\lambda_{\chi}/c_{\chi}}$, within which thermal effects simultaneously preserve the broken phase of $\chi$ and stabilize $\sigma$. The underlying mechanism is finite-temperature symmetry restoration \cite{Dolan:1973qd,Weinberg:1974hy,2013JCAP...01..017M,2013JHEP...12..053M}.

We restrict our attention throughout to those combinations of $n$ and $Q$ for which the vacuum displacement is negligible, that is, for which $\chi_{\rm min}\simeq\nu_{\chi}$. Within it, the waterfall field may be replaced by its symmetry-breaking value; the hat contributes nothing further, and the potential driving the second phase reduces to the portal term alone
\begin{equation}
  V(\phi)=\lambda_{\phi}\phi^{4}e^{-Q\phi},
  \label{eq45}
\end{equation}
where \(\lambda_{\phi}=\gamma^{4}\nu_{\chi}^{4}\).

The second phase is therefore a warm inflationary model of quartic type, this kind studied in \cite{Berghaus:2025}, where the amplitude constraint implies that the observed amplitude requires \(\lambda\sim 10^{-15}-10^{-21}\) across the range of dissipation strengths, with the self-coupling many orders of magnitude below its cold-inflation value. What distinguishes the present construction is the exponential modulation $e^{-Q\phi}$ inherited from the first phase: the potential has no minimum, so the inflaton is never trapped and continues to roll along the exponential tail after inflation ends, as required by the quintessential scenario.

By using the equations (\ref{eq2}) and (\ref{eq3}) and assuming the slow-roll conditions $|\ddot\phi|\ll(3H+\Gamma_\phi)|\dot\phi|$, $\tfrac{1}{2}\dot\phi^{2}\ll V$, $\rho_{r}\ll V$, $|\dot\rho_{r}|\ll4H\rho_{r}$ and $|24\dot fH^{3}|\ll3M_{p}^{2}H^{2}$, with \(f=f(\phi,\chi=\nu_\chi)=f_{wi}e^{nQ\phi}\) and \(2f_{wi}=f_{in}\). The background equations of the warm phase are then given by
\begin{equation}
  3M_{p}^{2}H^{2}\simeq\lambda_{\phi}\,\phi^{4}e^{-Q\phi},
  \label{eq46}
\end{equation}
\begin{equation}
  4H\rho_{r}\simeq\Gamma_\phi\dot\phi^{2},
  \label{eq47}
\end{equation}
\begin{equation}
\dot\phi\simeq\frac{-V_{\rm eff}'(\phi)}{3H(1+Q_{wi})},
  \label{eq48}
\end{equation}
where \(V_{\rm eff}=V(\phi)-f(\phi)\mathcal{G}\) with \(\mathcal{G}\simeq24H^4\) in the slow roll regime, while the dissipative friction $\Gamma_\phi\dot\phi$ and the source $\Gamma_\phi\dot\phi^{2}$ are related by energy conservation, and the strength of the effect is measured by $Q_{\rm wi}=\Gamma_\phi/3H$. We note that the growth of inflaton fluctuations induced by their coupling to the radiation is itself damped by the shear viscosity of the bath \cite{BasteroGil:2011xd}, an effect that will prove essential.

In the low-temperature regime $m_{X} \gg T$, the heavy mediator field can be integrated out, leading to an effective dissipative coefficient of cubic form \cite{BasteroGil:2012cm}
\begin{equation}
  \Gamma_{\phi}\sim\frac{T^{3}}{m_{X}^{2}} ,
  \label{eq49}
\end{equation}
where the $T^{3}$ dependence originates from the phase space of the light decay products, while the suppression by $m_{X}^{-2}$ reflects the propagator of the virtual mediator. 

In the present framework, the precise structure of $\Gamma_{\phi}$ depends on the microscopic origin of the mediator mass. Two natural realizations arise. If the mediator mass is set by a fixed high-energy scale, such as the Planck mass (as expected for higher-dimensional operators suppressed by gravity), then $m_{X}=\mathrm{const}$ and the dissipation scales as $\Gamma_{\phi}\propto T^{3}/M_{p}^{2}$. Alternatively, if the mediator acquires its mass dynamically through a Yukawa-type coupling $g_{X}\phi X\bar{X}$, then $m_{X}\propto\phi$ and one obtains $\Gamma_{\phi}\propto T^{3}/\phi^{2}$.

Both mechanisms are generically present in our model, and we therefore adopt the combined parametrization \cite{Sahu:2025}
\begin{equation}
  \Gamma_{\phi}(\phi,T)
  = \frac{c_{1}}{M_{p}^{2}} T^{3} + \frac{c_{2}}{\phi^{2}} T^{3},
  \label{eq50}
\end{equation}
where the coefficients $c_{1}$ and $c_{2}$ encode the underlying couplings and multiplicities of the mediator and light sectors. This form is not merely phenomenological and the two contributions dominate in different regimes. At large field values, the Planck-suppressed term is dominant, while as $\phi$ decreases toward the end of inflation, the inflaton-induced contribution becomes increasingly important.

A key consequence of the cubic temperature dependence is its effect on fluctuation dynamics. Since $\Gamma_{\phi}$ increases rapidly with $T$, a positive fluctuation in the radiation density enhances dissipation, which in turn feeds back into further radiation production. This positive feedback amplifies inflaton fluctuations and leads to a much stronger enhancement of scalar perturbations than in the linear dissipation case.

Having established the background equations, we turn to now the perturbations. The logic differs from the cold case in one essential respect; the inflaton fluctuation is not a vacuum mode stretched beyond the horizon, but a classical field driven by the thermal bath and damped by its own dissipation into that bath. Its amplitude is fixed not by the uncertainty principle but by the fluctuation-dissipation theorem \cite{PhysRevLett.75.3218,BERERA2000666}, and the scale at which it freezes is set by the friction rather than by the Hubble rate. Before deriving it, however, it is convenient to record the slow-roll parameters in the form the warm Gauss-Bonnet system requires. The potential parameters retain their usual definitions
\begin{equation}
  \epsilon_{V}=\frac{M_{p}^{2}}{2}\left(\frac{V'}{V}\right)^{2} ,
  \qquad
  \eta_{V}=M_{p}^{2}\frac{V''}{V} ,
  \qquad
  \beta_{\Gamma}=M_{p}^{2}\frac{\Gamma_{\phi}'V'}{\Gamma_{\phi} V} ,
  \label{eq51}
\end{equation}
but the conditions they must satisfy are relaxed, since the effective friction is $3H(1+Q_{\rm wi})$ and the effective slope is $V'(1-\lambda_{\mathcal{G}})$; one requires $\epsilon_{V}(1-\lambda_{\mathcal{G}})\ll1+Q_{\rm wi}$ and similarly for $\eta_{V}$ and $\beta_{\Gamma}$. It is this relaxation, by the factor $1+Q_{\rm wi}$, that admits potentials far too steep to inflate in vacuum. The first Hubble flow parameter follows from the slow-roll form of the inflaton equation as follows
\begin{equation}
  \epsilon_{1}=-\frac{\dot H}{H^{2}}
  \simeq\frac{\epsilon_{V}\bigl(1-\lambda_{\mathcal{G}}\bigr)}{1+Q_{\rm wi}},
  \label{eq52}
\end{equation}
where \(\lambda_{\mathcal{G}}=8f'V^{2}/3M_{p}^{4}V'\), the prime denotes the derivative over the inflaton field \(\phi\). Inflation ends when $\epsilon_{1}=1$.

Perturbing $\phi\to\phi+\delta\phi$ in the spatially flat gauge Eq.(\ref{eq17}), where the metric perturbations decouple at leading order in slow roll, the linearised equation of motion reads

\begin{equation}
  \delta\ddot\phi_{k}+\left(3H+\Gamma_{\phi}\right)\delta\dot\phi_{k}+\left(\frac{k^{2}}{a^{2}}+V''-f''\mathcal{G}\right)\delta\phi_{k}-f'\delta\mathcal{G}=\xi_{k}(t) .
  \label{eq53}
\end{equation}

The left-hand side of \eqref{eq53} is a damped oscillator; the difference between the cold case and the warm inflation case is the stochastic source on the right. This is not an arbitrary addition. Because the inflaton loses energy to the bath at the rate $\Gamma_{\phi}\dot\phi^{2}$, the same interaction necessarily returns energy in the form of random kicks, and the two are locked together by the fluctuation--dissipation theorem \cite{BERERA2000666,Gleiser:1993ea}: for a bath in equilibrium at temperature $T$,
\begin{equation}
  \bigl\langle\xi_{k}(t)\,\xi_{k'}(t')\bigr\rangle=2\,\Gamma_{\phi}\,T\,a^{-3}\,(2\pi)^{3}\delta^{3}(\mathbf{k}+\mathbf{k}')\,\delta(t-t') .
  \label{eq54}
\end{equation}

The essential feature is that the noise amplitude is proportional to the same $\Gamma_{\phi}$ that appears as friction. In the limit $\Gamma_{\phi}\to0$, the source vanishes; \eqref{eq53} reduces to the Mukhanov--Sasaki equation, and the familiar vacuum mode is recovered; in the opposite limit, the vacuum contribution is negligible, and the fluctuation is entirely thermal.

The scale at which the fluctuation freezes is correspondingly modified. In cold inflation, a mode ceases to evolve when $k=aH$, which is when Hubble friction overcomes the gradient term. Here the friction is larger by the factor $1+Q_{\rm wi}$, so freezing occurs earlier, at a physical wavelength still well inside the horizon. Neglecting $\delta\ddot\phi$, which is legitimate for $Q_{\rm wi}\gtrsim1$, equation \eqref{eq53} becomes
\begin{equation}
  3H\bigl(1+Q_{\rm wi}\bigr)\delta\dot\phi_{k}+\frac{k^{2}}{a^{2}}\delta\phi_{k}=\xi_{k} ,
  \label{eq55}
\end{equation}
and the mode relaxes towards its equilibrium value at the rate $\tau_{\rm rel}^{-1}=(k^{2}/a^{2})/3H(1+Q_{\rm wi})$. It can continue to respond to the noise only while this exceeds the expansion rate; setting $\tau_{\rm rel}^{-1}=H$ defines the freeze-out momentum
\begin{equation}
  \frac{k_{F}}{a}=H\sqrt{3(1+Q_{\rm wi})} ,
  \label{eq56}
\end{equation}
which for the dissipation strengths of interest lies several times above the horizon scale, and it is here, not at $k=aH$, that the amplitude must be evaluated \cite{Berera:2008ar}.

That amplitude follows from the stationary two-point function of \eqref{eq55}. For an overdamped Langevin equation $\gamma\dot x+\kappa x=\xi$ with $\langle\xi\xi\rangle=2D\,\delta(t-t')$ one has $\langle x^{2}\rangle=D/\gamma\kappa$; identifying $\gamma=3H(1+Q_{\rm wi})$, $\kappa=k^{2}/a^{2}$ and $D=\Gamma_{\phi} Ta^{-3}$ and forming the dimensionless spectrum $\mathcal{P}_{\delta\phi}=(k^{3}/2\pi^{2})\langle|\delta\phi_{k}|^{2}\rangle$
at $k=k_{F}$ gives, using $\Gamma_{\phi}=3HQ_{\rm wi}$,
\begin{equation}
  \mathcal{P}_{\delta\phi}=\frac{\sqrt{3}}{2\pi^{2}}HT\,\sqrt{Q_{\rm wi}},
  \label{eq57}
\end{equation}
the characteristic strong-dissipation scaling \cite{Hall:2003zp}. It is worth pausing on the combination $HT$ rather than $H^{2}$: with $T\gg H$, the thermal amplitude greatly exceeds the vacuum one, which is precisely why warm inflation accommodates a much steeper potential at the same observed amplitude.

Working in the spatially flat gauge $\psi=0$, where equation \eqref{eq18} simplifies, the curvature perturbation on flat slices is expressed as $\mathcal{R}=H\delta\phi/\dot\phi$. At leading order in the slow-roll approximation, metric perturbations decouple, as in the cold inflation scenario, provided that the radiation perturbation remains adiabatic on super-horizon scales. This requirement is met when the same field continuously sources the radiation bath. By substituting the slow-roll velocity equation (\ref{eq48}) and equation \eqref{eq57}, the primordial power spectrum for warm inflation is obtained, a result that has been extensively studied \cite{Hall:2003zp,2013JCAP...03..032R,2009JCAP...07..013G}
\begin{equation}
  \mathcal{P}_{\mathcal{R}}=\frac{V\left(1+Q_{\rm wi}\right)^{2}}{24\pi^{2}M_{p}^{4}\epsilon_{V} \left(1-\lambda_{\mathcal{G}}\right)^{2}}FG(Q_{\rm wi}) ,
  \label{eq58}
\end{equation}

where two factors \(G\) and \(F\), have been introduced to account for effects suppressed in the derivation above.

The first, $F$, collects the thermal statistics of the fluctuation. The inflaton fluctuation need not be in equilibrium with the bath, and its occupation number at freeze-out, $n_{*}=\bigl[\exp(k_{F}/aT)-1\bigr]^{-1}$, enhances the amplitude by $1+2n_{*}$ relative to the vacuum result; this factor reduces to unity for a mode in its ground state and to $2T/H\sqrt{3(1+Q_{\rm wi})}$ for one fully thermalised at the freeze-out scale \(k_F\). The sharp cutoff assumed in \eqref{eq56} must in addition be replaced by a smooth momentum profile, which introduces a $Q_{\rm wi}$-dependent coefficient. Carrying both through gives \cite{Hall:2003zp,2013JCAP...03..032R}
\begin{equation}
  F=1+2n_{*}+\frac{2\sqrt{3}\,\pi\,Q_{\rm wi}}
                  {\sqrt{3+4\pi Q_{\rm wi}}}\,\frac{T}{H} ,
  \label{eq59}
\end{equation}

whose last term dominates in our regime, where $T/H\sim10^{2}$ and $Q_{\rm wi}\gtrsim1$; for $Q_{\rm wi}\gg1$ it grows as $\sqrt{3\pi Q_{\rm wi}}\,T/H$, so that the thermal enhancement of the spectrum strengthens with the dissipation.

The second factor, $G(Q_{\rm wi})$, has a distinct physical origin and is responsible for making the temperature dependence of the dissipation coefficient observationally relevant. In deriving the power spectrum, we initially treated the radiation bath as a fixed external reservoir at temperature $T$. However, this approximation is not strictly valid: the radiation perturbation responds dynamically to the inflaton fluctuation $\delta\phi$ through the continuity equation. Since the dissipation coefficient depends on the temperature, $\Gamma_{\phi}\propto T^{c}$, this induces a feedback mechanism. 

In particular, a positive fluctuation in the radiation density increases the temperature, which enhances $\Gamma_{\phi}$, leading to stronger dissipation and further growth of the radiation perturbation. This feedback loop is therefore self-reinforcing, and its strength increases with the exponent $c$. Linearising the coupled inflaton–radiation system yields a set of perturbation equations whose growing mode amplifies the curvature perturbation between freeze-out and horizon crossing. This enhancement is conventionally encoded in the multiplicative factor $G(Q_{\rm wi})$ appearing in Eq.~\eqref{eq58}.

No closed-form analytical solution for this coupled system is known, and the function $G(Q_{\rm wi})$ must be determined numerically by solving the perturbation equations and fitting the result as a function of the dissipation ratio $Q_{\rm wi}$. Importantly, the resulting fit depends sensitively on the temperature dependence of \(\propto T^c\), for a linear dependence $c=1$, the amplification is relatively mild, whereas for the cubic dependence $c=3$ adopted in this work, the enhancement is significantly stronger. The two cases therefore differ quantitatively and cannot be interchanged.

For the cubic dissipation coefficient $\Gamma_{\phi}\propto T^{3}$, we adopt the parametrization \cite{Benetti:2016jhf}
\begin{equation}
  G(Q_{\rm wi}) = 1 + 4.981 Q_{\rm wi}^{1.946} + 0.127 Q_{\rm wi}^{4.330},
  \label{eq60}
\end{equation}
which provides an accurate fit over the range of dissipation strengths relevant to our analysis. We further define the logarithmic slope by the following expression 
\begin{equation}
  \beta_{G} \equiv \frac{d\ln G}{d\ln Q_{\rm wi}},
\end{equation}
this characterizes the amplification's sensitivity to dissipation strength. 
 
Treating the radiation as a perfect fluid and a bath continuously driven out of equilibrium by dissipation is not the same. Departures from local equilibrium generate viscous pressures, of which the shear component is the relevant one here and opposes the anisotropic stresses that develop as the radiation responds to the inflaton fluctuation, and therefore acts as an additional friction on precisely the perturbations that sustain the growing mode \cite{BasteroGil:2011xd,Weinberg:1971mx}. The strength of the effect is measured by the dimensionless shear viscosity $\bar\zeta_{s}=\zeta_{s}H/3\rho_{r}$, which compares the viscous relaxation rate to the expansion rate, and its influence on the spectrum is captured to good accuracy by \cite{BasteroGil:2011xd}
\begin{equation}
  F_{s}\bigl(\bar\zeta_{s}\bigr)=\frac{1}{2}
  \Bigl[1-\tanh\bigl(\log_{10}\bar\zeta_{s}+\Delta_{s}\bigr)\Bigr],
  \label{eq62}
\end{equation}
where the function \(G\) shift to be \(G_{\rm eff}=G^{F_{s}} \) and the logarithmic slop \(\beta_{G}\to F_{s}\frac{d\ln G}{d\ln Q_{\rm wi}}\) and $\Delta_{s}$ is a fitted constant locating the transition: $F_{s}=1/2$ at $\bar\zeta_{s}=10^{-\Delta_{s}}$, so that the damping sets in once the viscosity exceeds that threshold. It is obtained by matching \eqref{eq62} to the numerical solution of the perturbation equations including the shear viscous terms \cite{BasteroGil:2011xd}, and we adopt the value $\Delta_{s}=1.36$ appropriate to a quartic potential with cubic dissipation. Hence, for $\bar\zeta_{s}\gg1$ the growing mode is removed entirely, and for $\bar\zeta_{s}\ll1$ the undamped result is recovered. Since $\zeta_{s}\propto T^{3}$ in a weakly coupled bath \cite{Arnold:2000dr}, one has $\bar\zeta_{s}\propto H/T$, so that the damping is most effective exactly where the thermal enhancement is largest, and the two effects are not independent.

The spectral index now follows by differentiation. On super-horizon scales $d\ln k=dN=H\,dt$, and since every background quantity depends on time only through $\phi$ one has $d/d\ln k=(\dot\phi/H)\,d/d\phi$. Taking the logarithm of \eqref{eq58} along with $U_{\mathcal{G}}=V'(1-\lambda_{\mathcal{G}})$, one arrives at
\begin{equation}
  n_{s}=1-\left(\frac{5}{2}-\beta_{G}\right)\epsilon_{1}+\frac{U_{\mathcal{G}}}{H\Gamma_{\phi}}\left[2\frac{U_{\mathcal{G}}'}{U_{\mathcal{G}}}-\left(\frac{5}{2}+ \beta_{G}\right)\frac{\Gamma_{\phi}'}{\Gamma_{\phi}}-\frac{T'}{T}\right] ,
  \label{eq63}
\end{equation}
where primes denote $d/d\phi$, the structure of this expression repays reading. The first term in Eq. (\ref{eq63}) is the familiar slow-roll suppression, modified by $\beta_{G}$; the growing mode contributes with the opposite sign to $\epsilon_{1}$ and therefore tilts the spectrum blue, which is why the shear damping of \eqref{eq62} is indispensable rather than merely quantitative. The bracket collects the running of the dissipation and the temperature, each weighted by its own power; these terms have no counterpart in cold inflation and allow a quartic potential, excluded in vacuum, to remain viable here. In practice, we evaluate the derivatives numerically from the background solution rather than expanding them analytically, since $T(\phi)$ is obtained by solving the attractor condition in equation (\ref{eq47}) and admits no closed form.

Tensor modes, finally, are not sourced by the bath at leading order, because the dissipative interaction couples to $\delta\phi$ and not to the transverse-traceless metric perturbation; the tensor spectrum therefore carries no thermal enhancement. It does, however, retain the Gauss-Bonnet correction, since the higher-curvature term modifies the quadratic action for the tensor modes,
\begin{equation}
  \mathcal{P}_{t}=\frac{2H^{2}}{\pi^{2}M_{p}^{2}\left(1-\delta_1\right)c_T^2},
  \label{eq64}
\end{equation}

Forming the ratio \(r\) with \eqref{eq58} and \eqref{eq64} in the strong-dissipation limit,
\begin{equation}
  r=\frac{\mathcal{P}_{t}}{\mathcal{P}_{\mathcal{R}}}\simeq\frac{32\epsilon_{V}\left(1-\lambda_{\mathcal{G}}\right)^{2}}{\sqrt{3\pi}Q_{\rm wi}^{5/2}G_{\rm eff}\left(1-\delta_1\right)c_T^3} .
  \label{eq65}
\end{equation}

The steep suppression $r\propto Q_{\rm wi}^{-5/2}$ is the characteristic signature of the scenario where the thermal bath enhances the scalar amplitude while the tensor amplitude is not. However, that strong dissipation drives $r$ far below its cold-inflation value at the same $\epsilon_{V}$. It is precisely this that allows the quartic potential of the warm phase, for which the cold prediction $r\simeq16/N_{*}\simeq0.3$ is excluded by more than an order of magnitude, to satisfy the observational bound.

We now address the numerical solution of the previously derived system. As outlined at the beginning of this section, for each pair $(n, Q)$, viable parameter values are selected to ensure that $\chi_{\rm min} \simeq \nu_\chi$ and that the potential reduces to be $\propto \lambda_\phi \phi^4$. This is accomplished by determining $\gamma$ such that it aligns with the observed scalar amplitude $A_{s} \simeq 2.1 \times 10^{-9}$. Furthermore, the requirement that the total number of e-folds lies between 55 and 60 before the end of inflation, when the pivot scale exits the horizon at $k_* = 0.05 \text{ Mpc}^{-1}$, sets the value of $c_{1}$, while $c_{2}$ is fixed at $10^6$ \cite{BasteroGil:2012cm,Benetti:2016jhf}. Since $\phi_{c}$ depends on $n$ and $Q$ only through the combination $(n-1)Q$, and all derived quantities inherit this dependence, the parameter space is effectively one-dimensional. The analysis therefore begins by displaying the region of the $(n,Q)$ plane in which the construction remains consistent. 

\begin{figure}[H]
\centering
\includegraphics[width=0.6\textwidth]{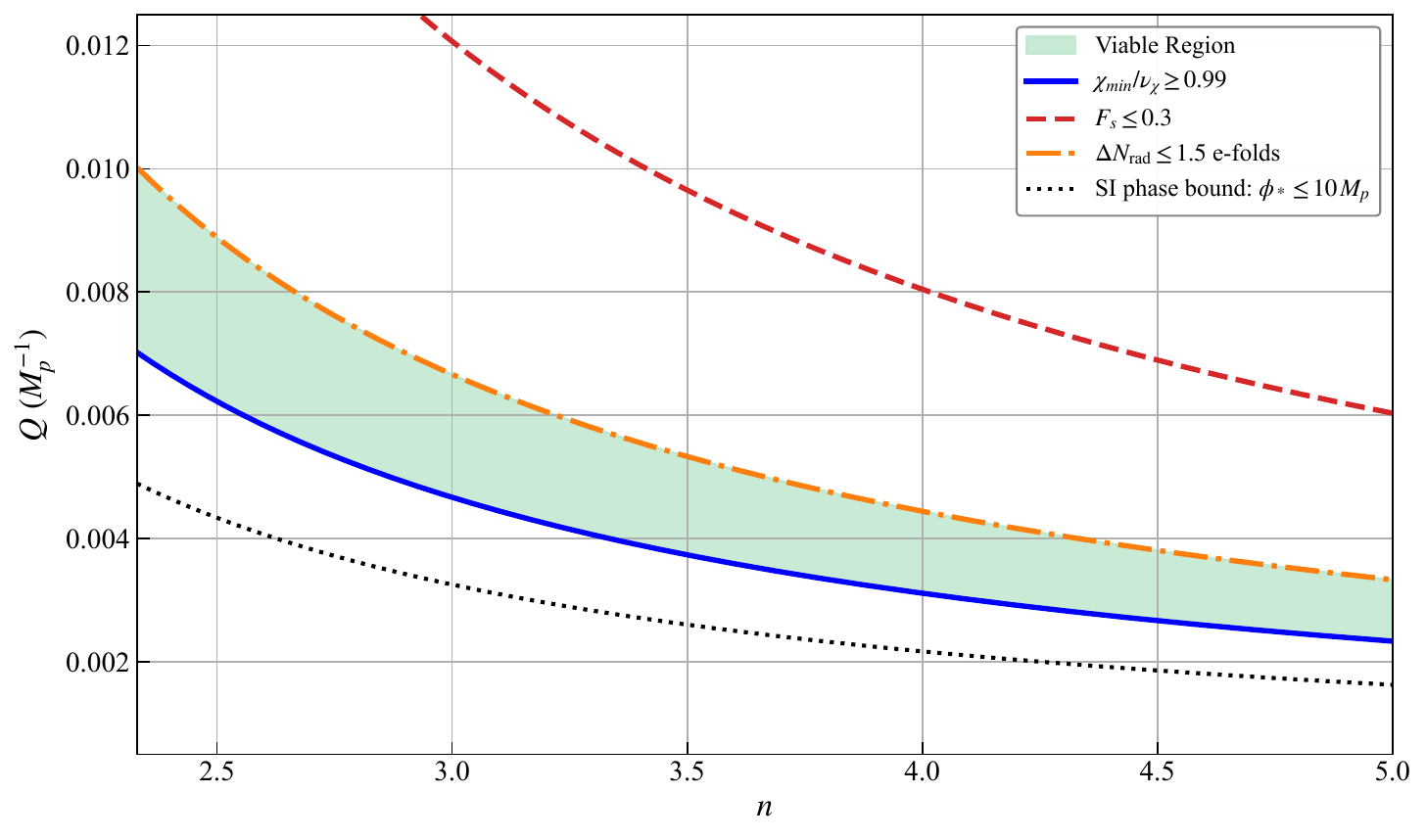}
\caption{Viable region in the $(n,Q)$ parameter space (green shaded band). Each constraint corresponds to a curve of constant $(n-1)Q$, so that the allowed domain forms a band bounded by two hyperbolae. The lower boundary (blue solid line) arises from the requirement that the portal interaction does not significantly displace the waterfall vacuum, $\chi_{\min} \geq 0.99\,\nu_{\chi}$. In contrast, the upper boundary (orange dash-dotted line) ensures that the post-waterfall radiation phase remains sufficiently short ($\Delta N_{\rm rad} \lesssim 1.5$), meaning the thermal bath temperature $T$ redshifts away rapidly enough and track the equilibrium. The condition $F_s \leq 0.3$ (red dashed line), which guarantees efficient shear damping, the super-inflationary phase bound $\phi_* \leq 10 M_p$ (black dotted line).}
\label{fig5}
\end{figure}

Figure~\ref{fig5} illustrates this region, which is bounded from below by the requirement that the portal does not appreciably displace the waterfall vacuum, $\chi_{\rm min} \simeq \nu_{\chi}$. Conversely, the parameter space is bounded from above by two crucial cosmological requirements: first, the shear viscous damping must remain effective ($F_s \leq 0.30$) to prevent the growing mode from dominating the perturbation spectrum, and second, the radiation phase triggered by the waterfall transition must dilute away rapidly ($\Delta N_{\rm rad} \leq 1.5$ e-folds) to prevent the premature re-entry of large-scale modes that would erase the low-$\ell$ primordial imprint. Numerically, we find the following bounds on $(n-1)Q$:
\begin{equation}
  9.34 \times 10^{-3}M_p^{-1} \leq (n-1)Q \leq 1.333 \times 10^{-2}M_p^{-1}.
  \label{eq66}
\end{equation}

The model does not select $n$ or $Q$ independently; instead, it specifies a relation between them. The requirement for effective shear damping, $F_{s}\lesssim0.3$, is automatically satisfied wherever the upper bound in \eqref{eq66} applies. Additionally, the Planck constraints on $n_{s}$ are not restrictive, as the observed tilt is reproduced over a range in $(n-1)Q$ that is approximately three times broader than the interval determined by internal criteria. Therefore, agreement with the spectral index is a general feature of this framework, while the prediction in \eqref{eq66} arises from the model's internal consistency.

After establishing the viable region in the \((n,Q)\) plane, it is informative to analyze the two quantities that determine both the strength and reliability of the dissipative description. Table~\ref{tab1} presents the ranges of the dissipation normalization \(c_1\) and the shear parameter \(F_s\) from equation (\ref{eq62}) at horizon crossing, evaluated across the same admissible window of \(Q\) in (\ref{eq66}), with the companion normalization held fixed at \(c_2=10^{6}\). Both quantities grow monotonically with \(Q\), \(c_1\) increasing by a factor of about \(3.4\) across the window while \(F_s\) remains confined to the interval \(0.05\lesssim F_s\lesssim 0.08\). The smallness of \(F_s\) throughout the viable region is significant; the shear correction to the scalar power spectrum stays at the few-percent level. It never becomes the limiting requirement, instead, the window boundaries are set by the \(\chi\) condition from below and \(\Delta N_{\rm rad}\) condition from above.

\begin{table}[H]
    \centering
    \caption{Ranges of the dissipation normalisation \(c_1\) and of the shear parameter \(F_s\) Eq.(\ref{eq62}) over the theoretically viable window of \(Q\) for each value of \(n\), evaluated at horizon crossing. The window is fixed by the equation (\ref{eq66}), and every entry is quoted as \(\mathrm{min}-\mathrm{max}\); both quantities increase monotonically with \(Q\), so the two ends of each interval correspond to the two edges of the window. The companion normalization is held fixed at \(c_2=10^{6}\) throughout the scan. The shear correction remains at the few-percent level over the entire viable region, well below the threshold at which it would become an active constraint.}
    \setlength{\tabcolsep}{7pt}      
    \setlength{\arrayrulewidth}{1pt} 
    \begin{tabular}{cccc}
        \hline
        \hline
        \(n\) & \(Q(10^{-3}M_p^{-1})\) & \(c_1(10^{6})\) & \(F_s\)\\
        \hline
        \(2.33\) & \(7.020-10.020\) & \(4.416-15.231\) & \(0.0504-0.0777\)\\
        \(3.00\) & \(4.668-6.663\) & \(4.435-15.309\) & \(0.0505-0.0780\)\\
        \(3.50\) & \(3.735-5.331\) & \(4.444-15.338\) & \(0.0506-0.0781\)\\
        \(4.00\) & \(3.112-4.442\) & \(4.444-15.355\) & \(0.0506-0.0781\)\\
        \hline
    \end{tabular}
    
    \label{tab1}
\end{table}

Table~\ref{tab2} summarizes the parameter ranges and resulting observables across the admissible band. For fixed \(n\), the table reports the window in \(Q\), the quartic coupling \(\lambda_\phi\), the dissipation ratio \(Q_{\rm wi}\), the temperature at horizon crossing (both in absolute terms and relative to the Hubble rate), the number of e-folds \(N_{*}\) derived from the pivot scale, and the predicted values of \(n_{s}\) and \(r\). The parameter \(\gamma\) varies only slightly, with the amplitude constraint limiting it to \((1.01-1.20)\times10^{-2}M_p^{-1}\), corresponding to a variation of only a few tens of percent across the range. The quartic coupling, determined by \(\lambda_\phi=\gamma^4\lambda_\chi^4\), remains within a factor of two of \(10^{-16}\). The observables are similarly constrained, with \(N_{*}\) restricted to \(56.3-57.0\) and \(n_{s}\) to \(0.961-0.963\). In contrast, the tensor-to-scalar ratio is the most sensitive observable, varying by a factor of \(5.3\) across the window and thus serving as the most effective observational discriminant between the two edges. Additionally, all four rows in the table are numerically nearly identical; the dependence on \(n\) is absorbed into the window location via the scaling \((n-1)Q=\mathrm{const}\), so the model effectively traces a single curve sampled at four points.
\begin{table}[H]
    \caption{Ranges of the model parameters and of the inflationary observables at horizon crossing, evaluated over the theoretically viable window of \(Q\) for each value of \(n\). The window is fixed by the single scaling relation (\ref{eq66}), so that the admissible interval in \(Q\) simply contracts as \((n-1)^{-1}\). Every entry is quoted as \(\mathrm{min}-\mathrm{max}\); all listed quantities vary monotonically with \(Q\), the lower end of the \(Q\) window corresponding to the largest values of \(\lambda_\phi\), \(T\), \(N_*\), \(n_s\) and \(r\), and to the smallest values of \(T/H\) and \(Q_{\rm wi}\). The parameter \(\gamma\) is not a free input: at each point of the scan it is fixed by matching the observed amplitude \(A_s\) of the scalar power spectrum at the pivot scale \(k_*=0.05\,\mathrm{Mpc}^{-1}\), and spans \(1.01\times10^{-2}-1.20\times10^{-2}\) over the whole viable region. The quartic coupling then follows as \(\lambda_\phi=\gamma^{4}\lambda_\chi^{4}\).}
    \setlength{\tabcolsep}{6pt}      
    \setlength{\arrayrulewidth}{1pt} 
    \small
    \hspace{-2cm}\begin{tabular}{ccccccccc}
        \hline
        \hline
        \(n\) & \(Q(10^{-3}M_p^{-1})\) & \(\lambda_\phi(10^{-16})\) & \(T(10^{14}\mathrm{GeV})\) & \(T/H\) & \(Q_{\rm wi}\) & \(N_*\) & \(n_s\) & \(r(10^{-3})\)\\
        \hline
        \(2.33\) & \(7.020-10.020\) & \(1.086-2.039\) & \(1.564-2.564\) & \(269.6-459.2\) & \(6.69-14.61\) & \(56.34-57.04\) & \(0.9616-0.9631\) & \(1.98-10.40\)\\
        \(3.00\) & \(4.668-6.663\) & \(1.055-1.986\) & \(1.561-2.561\) & \(270.5-461.0\) & \(6.73-14.68\) & \(56.34-57.02\) & \(0.9614-0.9628\) & \(1.97-10.33\)\\
        \(3.50\) & \(3.735-5.331\) & \(1.043-1.966\) & \(1.560-2.559\) & \(270.9-461.7\) & \(6.74-14.71\) & \(56.34-57.03\) & \(0.9613-0.9627\) & \(1.96-10.30\)\\
        \(4.00\) & \(3.112-4.442\) & \(1.035-1.952\) & \(1.559-2.559\) & \(271.0-462.1\) & \(6.74-14.72\) & \(56.34-57.02\) & \(0.9613-0.9627\) & \(1.96-10.30\)\\
        \hline
    \end{tabular}
    
    \label{tab2}
\end{table}

\begin{figure}[H]
\centering
\begin{tabular}{c c}
    \includegraphics[width=0.52\textwidth]{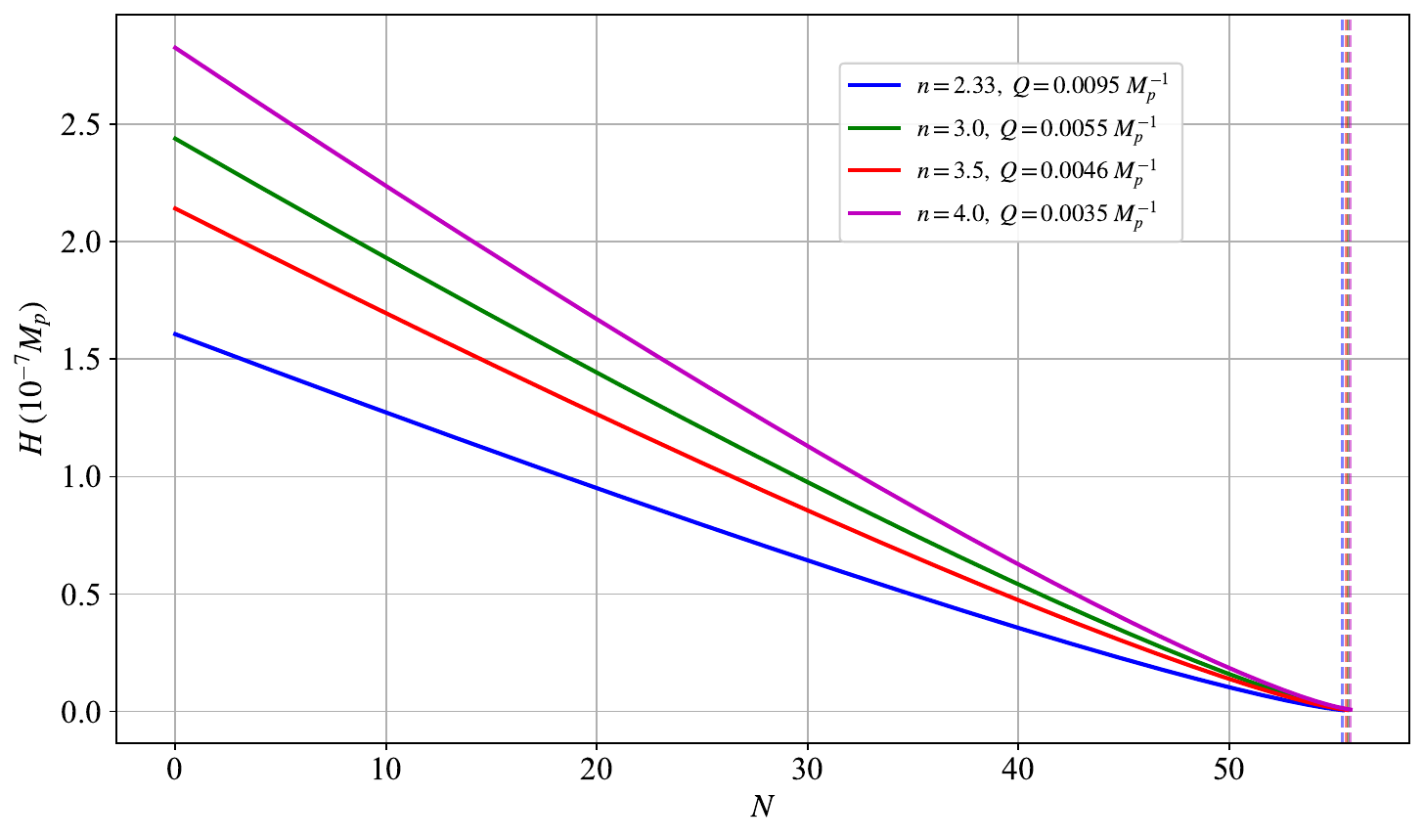} &
    \includegraphics[width=0.52\textwidth]{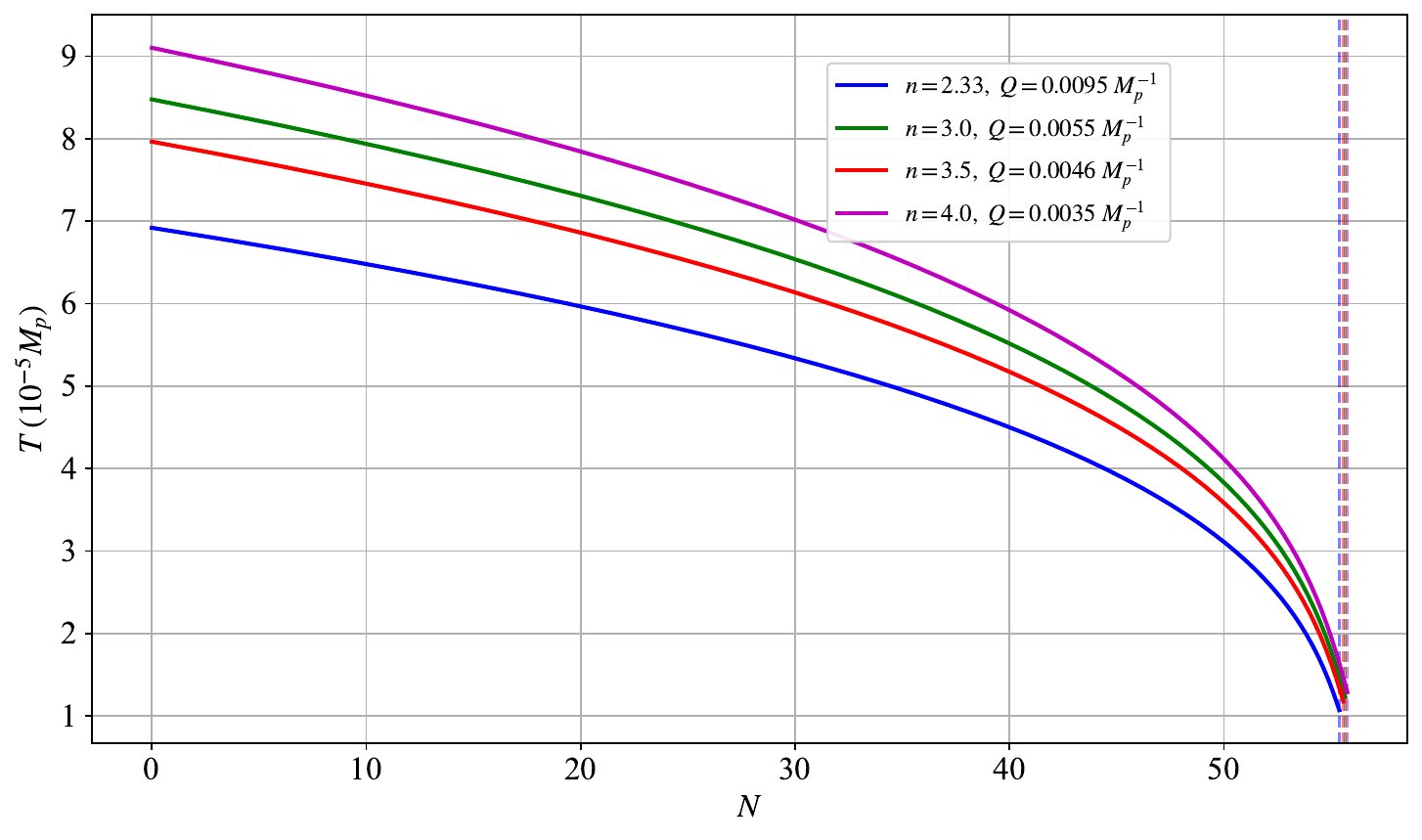}\\
    \includegraphics[width=0.52\textwidth]{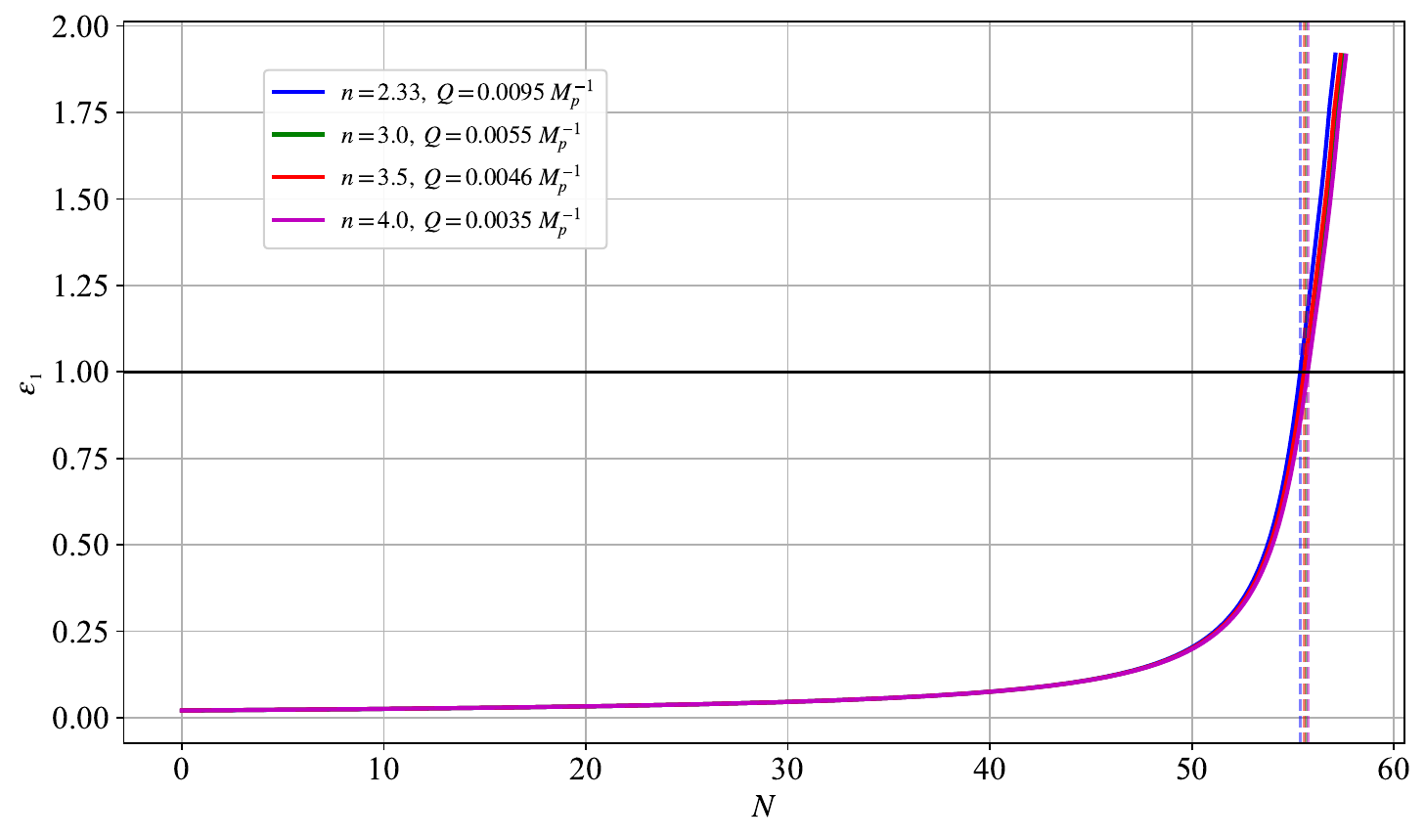} &
    \includegraphics[width=0.52\textwidth]{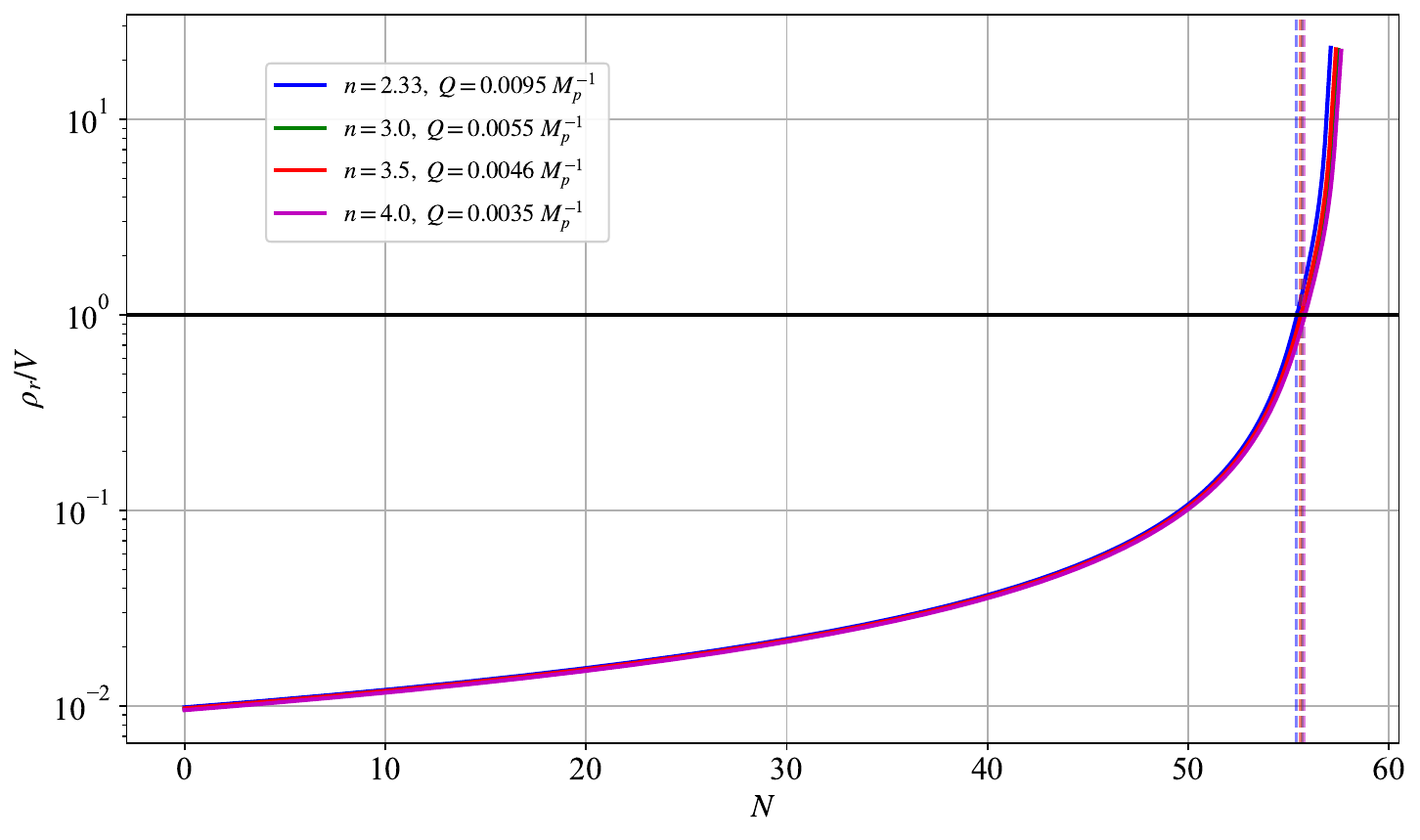}
\end{tabular}
\caption{Background evolution of the warm phase for the parameter sets of
Table~\ref{tab1} and Table~\ref{tab2}, as functions of the number of e-folds measured from the
onset of warm inflation. Upper left: the Hubble rate. Upper right: the
temperature of the radiation bath. Lower left: the first slow-roll parameter
$\epsilon_{1}$, which remains small until the end of inflation at
$\epsilon_{1}=1$. Lower right: the ratio $\rho_{r}/V$, which settles at the
per cent level, the expansion remaining potential dominated while the
radiation is continuously replenished.}
\label{fig6}
\end{figure}

Figure~\ref{fig6} shows the background evolution of the four representative cases, with \(N\) measured from horizon crossing. In every case, inflation ends after \(55 - 60\) e-folds, marked by the dashed vertical lines at which the first slow-roll parameter reaches unity, in agreement with the values of
\(N_{*}\) reported in Table~\ref{tab2}. The evolution of \(\epsilon_{1}\) is that of a well-behaved slow-roll trajectory: starting from \(\epsilon_{1}\simeq0.02\) at horizon crossing, it remains below \(0.1\) for the first fifty e-folds and rises steeply only in the last e-folds, so that the observable window is generated deep in the slow-roll regime and the approach to the end of inflation is abrupt but smooth. The Hubble rate declines almost linearly from \((1.6-2.8)\times10^{-7}M_p\), falling by more than an order of magnitude over the same interval, while the temperature decreases monotonically from \((6.9-9.1)\times10^{-5}M_p\) to about \(1.2\times10^{-5}M_p\), that is by a factor of roughly seven. Because \(T\) falls considerably more slowly than \(H\), the ratio \(T/H\) grows steadily from its value of a few hundred at horizon crossing. The condition \(T\gg H\) that defines warm inflation is satisfied by a wide margin not only at the pivot scale but throughout the entire inflationary era. The lower right panel makes the same point in terms of energy densities: the radiation fraction \(\rho_{r}/V\) grows smoothly and without feature from \(10^{-2}\) at horizon crossing to \(10^{-1}\) around \(N\simeq50\), and crosses unity at precisely the e-fold where \(\epsilon_{1}=1\). Inflation is thus potential-dominated over the whole observable range while sustaining a radiation bath that is never negligible, and it terminates by handing over to radiation domination continuously; the same dissipative coupling that maintains the bath during inflation delivers the transition to a radiation-dominated universe at its end. Finally, the two panels showing \(\epsilon_{1}\) and \(\rho_{r}/V\) are almost perfectly degenerate across the four cases, whereas \(H\) and \(T\) separate cleanly; the ordering of the latter is governed not by \(n\) but by the combination \((n-1)Q\), equal to \(1.264,\,1.150,\,1.100\) and \(1.050\times10^{-2}M_p^{-1}\) for the cases displayed, with the largest value lying lowest in both panels. The dimensionless dynamics is therefore common across the entire viable band.

\begin{figure}[H]
\centering
\begin{tabular}{c c}
    \includegraphics[width=0.52\textwidth]{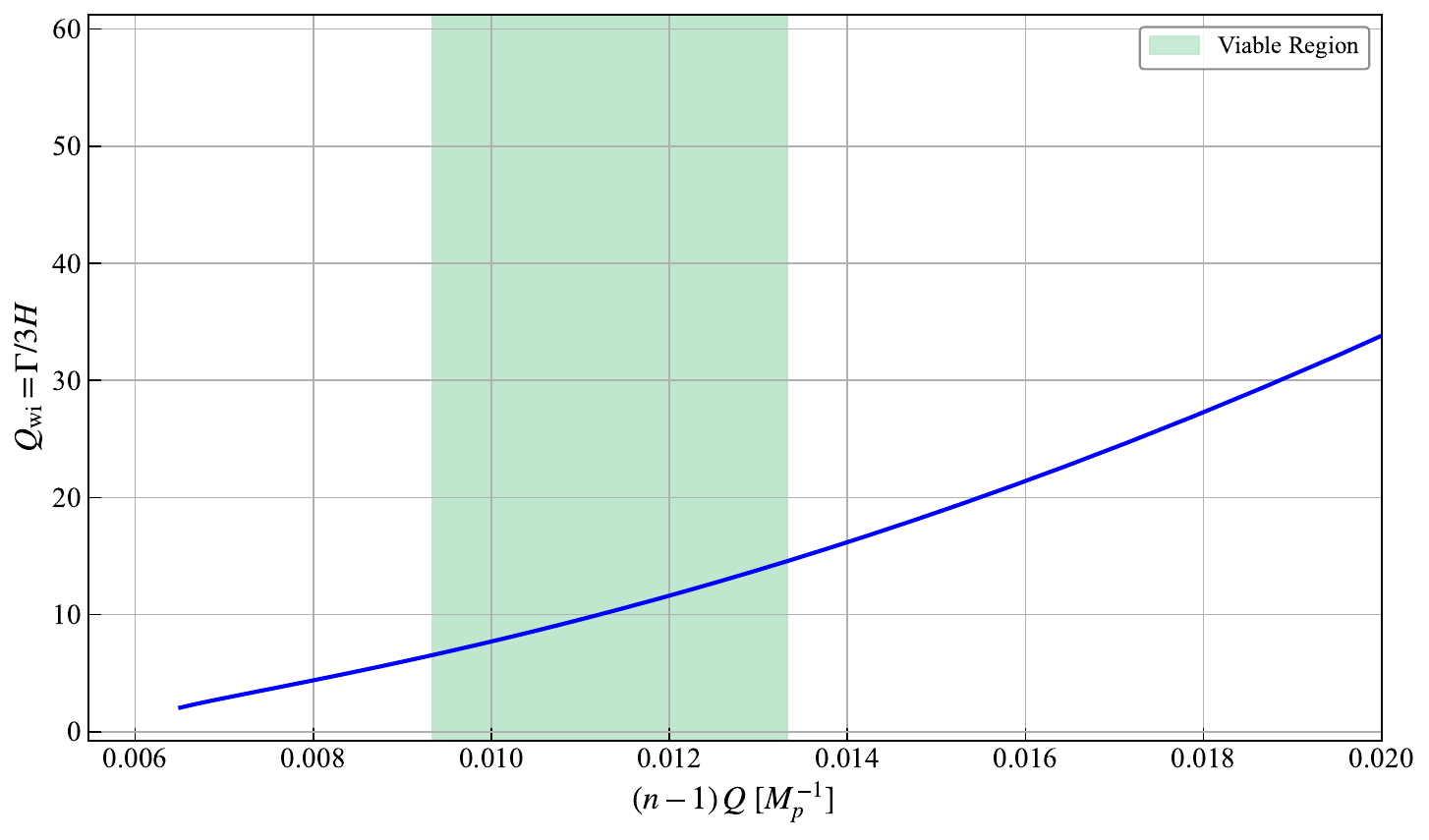} &
    \includegraphics[width=0.52\textwidth]{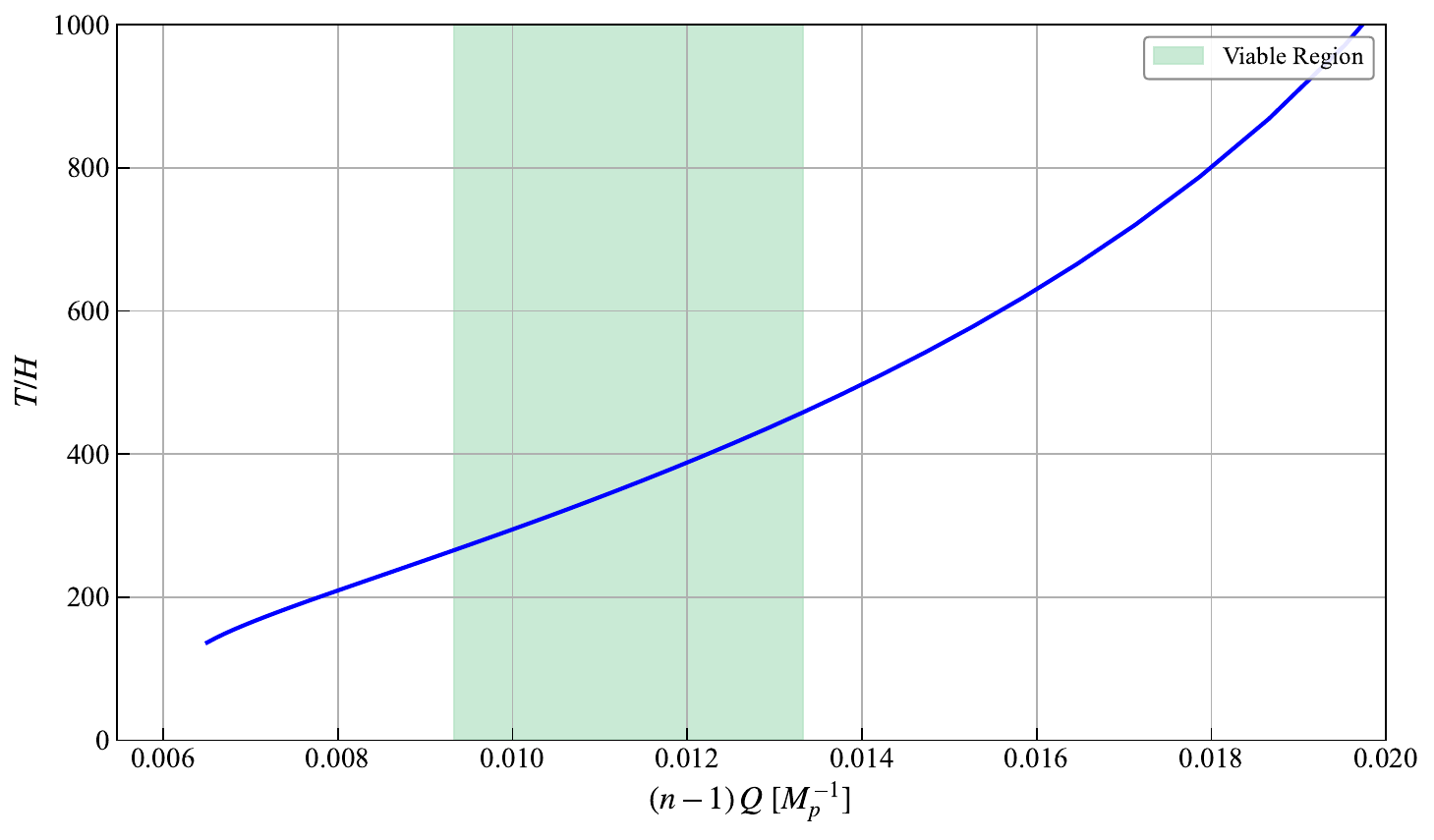}
\end{tabular}
\caption{The dissipation ratio \(Q_{\rm wi}=\Gamma_{\phi}/3H\) (left) and the ratio \(T/H\) (right), both
evaluated at horizon crossing, as functions of \((n-1)Q\); the shaded region is the viable band in Eq.(\ref{eq66}). Cases with different \(n\) and \(Q\) but equal
\((n-1)Q\) fall on the same curve, confirming that the parameter space is effectively one dimensional. Within
the band \(Q_{\rm wi}=6.5-14.6\), placing the model in the strongly dissipative regime, and \(T/H=266-458\),
ensuring that the fluctuations are thermal rather than quantum in origin.}
\label{fig7}
\end{figure}

\begin{figure}[H]
\centering
\includegraphics[width=0.6\textwidth]{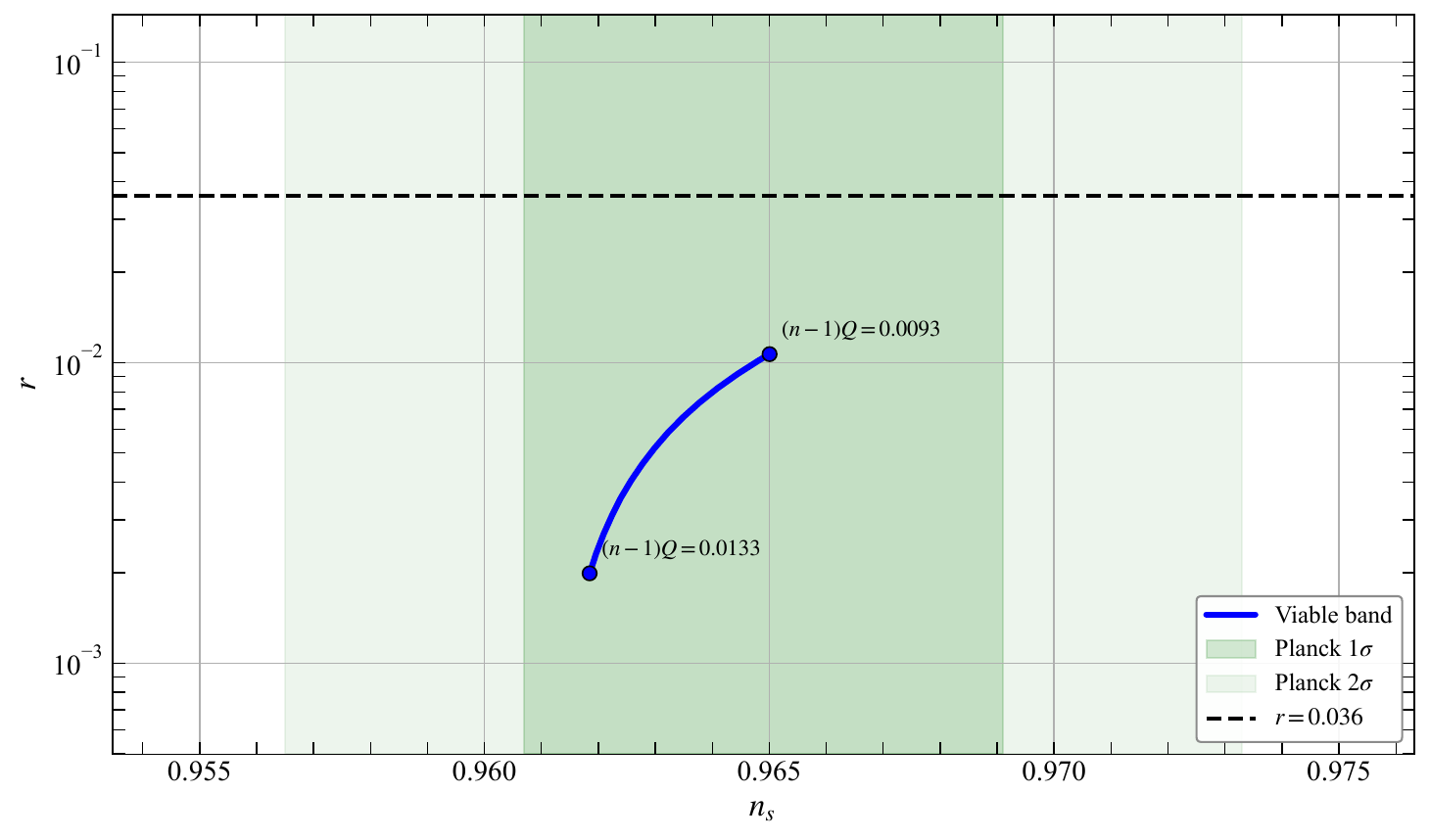}
\caption{Predictions in the \((n_{s},r)\) plane across the viable band, shown against the Planck and BICEP/Keck constraints \cite{Planck2018X,PhysRevLett.127.151301}. The band is traced by varying \((n-1)Q\) between the two limits indicated, \(r\) decreasing as the dissipation strengthens.}
\label{fig8}
\end{figure}

Figure~\ref{fig7} shows two key quantities that measure the effect at horizon crossing, the dissipation ratio \(Q_{\rm wi}\) and the ratio \(T/H\), both plotted against \((n-1)Q\). Each is a single-valued function of this combination, which confirms the expected degeneracy. This means cases with different values of \(n\) and \(Q\) but the same product all fall on the same curve. Throughout the viable range, dissipation remains strong, \(Q_{\rm wi}\) increases from \(6.5\) at the lower end to \(14.6\) at the upper end. This puts the model clearly in the regime where the thermal bath, not the vacuum, sets the fluctuation amplitude. Over the same range, \(T/H\) rises from \(266\) to \(458\), which is more than two orders of magnitude above one. This means the warm description is not just valid, but comfortably so.
Figure~\ref{fig8} displays the predicted observables in the \((n_{s},r)\) plane. The spectral index spans from \(0.9619\) to \(0.9650\) across the viable band, yielding a spread of \(3.2\times10^{-3}\), which is smaller than the Planck uncertainty \(\sigma(n_s)=4.2\times10^{-3}\). Consequently, the prediction appears as a point rather than a curve in this direction, and the entire band is contained within the \(1\sigma\) contour. The lower edge coincides with the central value \(n_s=0.9649\). In contrast, the tensor-to-scalar ratio varies by a factor of \(5.4\), from \(1.07\times10^{-2}\) at the lower edge to \(1.99\times10^{-3}\) at the upper edge, decreasing with dissipation strength according to \(r\propto Q_{\rm wi}^{-5/2}\). Even at its maximum, the prediction remains three times below the current bound \(r<0.036\), and the entire band lies within the anticipated sensitivity of forthcoming polarisation experiments. This renders the model testable in a manner that the spectral index alone does not permit. If primordial tensors are detected within this range, together with the measured tilt, it would determine \((n-1)Q\) and thereby constrain the entire model.

\section{Post-inflationary epoch}
\label{sec5}

Warm inflation ends when $\epsilon_{1}$ reaches unity, but the inflaton is not thereby removed from the dynamics. The potential has no minimum, so $\phi$ continues to roll, and its energy density falls monotonically, leaving behind the radiation generated by dissipation. The second scalar of the model, $\sigma$, has so far played no dynamical role, having been held at the origin throughout; it is its eventual release that constitutes the second essential ingredient of the construction, and the purpose of this section is to trace it.

Two distinct mechanisms maintain $\sigma$ at $\sigma\simeq0$, operating sequentially. The first mechanism arises from the higher-curvature term. The coupling function includes $-\sigma^{2}/\Lambda^{2}$, which, through $V_{\rm eff}=V-f\mathcal{G}$, produces a positive contribution $m_{\sigma,\mathcal{G}}^{2}\propto\mathcal{G}/\Lambda^{2}$ to the transverse curvature. Since $\mathcal{G}=24H^{2}(\dot H+H^{2})$, this contribution scales as $H^{4}$ and becomes significant during inflation, but it diminishes rapidly as the expansion rate decreases. The second mechanism becomes dominant as the first subsides. After the dissipatively generated bath reaches equilibrium, the coupling of $\sigma$ to the light sector induces the well-known thermal correction $c_{\sigma}T^{2}\sigma^{2}$ \cite{Dolan:1973qd,Weinberg:1974hy}. Because $T$ decreases only as $a^{-1}$, while $\mathcal{G}$ decreases as $a^{-8}$ during radiation domination, the thermal term quickly overtakes the geometric contribution after inflation and continues to stabilize $\sigma$ at the origin long after the geometric effect becomes negligible. The effective mass of the field \(\sigma\) at the false vacuum is given by 
\begin{equation}
    m_{\sigma\simeq0}^2=c_\sigma T^{2}-\lambda_{\sigma}\nu_{\sigma}^{2}e^{-Q\phi}.
\end{equation}

The symmetry is broken when the thermal mass can no longer overcome the tachyonic one, that is at \(T_{c}=\nu_{\sigma}\sqrt{\lambda_{\sigma}/c_{\sigma}}e^{\frac{-Q\phi_{_T}}{2}}\). 

A potential concern is that $\sigma$ may not be stable at $\nu_{\sigma}$, as the portal term includes the factor $(1-\sigma^{4}/\nu_{\sigma}^{4})$, which contributes to the transverse force. Differentiating the potential in the $\sigma$ direction,
\begin{equation}
  \frac{\partial V}{\partial\sigma}=\left[\lambda_{\sigma}\sigma\bigl(\sigma^{2}-\nu_{\sigma}^{2}\bigr)-\frac{4\lambda_\phi\phi^{4}\sigma^{3}}{\nu_{\sigma}^{4}}\right]e^{-Q\phi}.
  \label{eq68}
\end{equation}

At $\sigma=\nu_{\sigma}$, the expression does not vanish but instead yields $-4\lambda_\phi\phi^{4}e^{-Q\phi}/\nu_{\sigma}$. Consequently, the symmetry-breaking scale does not correspond to an extremum while the portal remains active, resulting in the true minimum being shifted to $\sigma_{\rm min}=\nu_{\sigma}(1+\delta_\sigma)$, where \(\delta_\sigma=2\lambda_\phi\phi^{4}/\lambda_{\sigma}\nu_{\sigma}^{4}\). The curvature is given by
\begin{equation}
   \frac{\partial^{2}V}{\partial\sigma^{2}}\bigg|_{\nu_{\sigma}}=2\lambda_{\sigma}\nu_{\sigma}^{2}e^{-Q\phi}\left[1-\frac{6\lambda_\phi\phi^{4}}{\lambda_{\sigma}\nu_{\sigma}^{4}}\right].
  \label{eq69}
\end{equation}

The expression is positive only when $\lambda_{\sigma}\nu_{\sigma}^{4}>6\lambda_\phi\phi^{4}$. If the portal remains at an inflationary magnitude, the transverse direction becomes tachyonic, and $\sigma$ diverges instead of stabilizing.

However, the inflaton has progressed significantly along the exponential tail, causing the potential $\propto\phi^{4}e^{-Q\phi}$ to decrease by many orders of magnitude relative to its value at horizon crossing and approach zero. In contrast, the bare potential of $\sigma$ is determined solely by $\nu_{\sigma}$. Setting $\lambda_{\sigma}=10^{-2}$ and $\nu_{\sigma}=0.8$ eV at $z=3406$ yields a potential magnitude of approximately $1.0\times10^{-3}$ eV$^{4}$. Stability therefore only requires that the $\gamma^4$-potential remains below $6.8\times10^{-4}$ eV$^{4}$ at the transition. Both conditions are satisfied by a substantial margin, and the portal interaction becomes negligible well before the temperature decreases to an electronvolt. Consequently, the $\sigma$ transition is governed exclusively by the Mexican hat potential and the thermal correction, allowing $\sigma$ to settle at $\nu_{\sigma}$ without competition.

The consequence of the transition for the higher-curvature sector is
immediate, and it is fixed at the moment $\sigma$ first reaches $\nu_{\sigma}$, before any oscillation has taken place. Evaluating the coupling function at the two vacuum states then occupied, $\chi=\nu_{\chi}$ and $\sigma=\nu_{\sigma}$, the quartic term $\sigma^{4}/M_{s}^{4}$ now dominates, and $f=f_{0}e^{nQ\phi}$ with
\begin{equation}
  f_{0}\simeq2f_{1}g^{4}\nu_{\chi}^{4}\left(\frac{\nu_{\sigma}}{M_{s}}\right)^{4}=2f_{in}\left(\frac{\nu_{\sigma}}{M_{s}}\right)^4.
  \label{eq70}
\end{equation}

Since $f_{0}$ is not a fundamental constant but an effective parameter of the model, its value cannot be determined a priori and must instead be fixed by matching to phenomenological requirements. In particular, requiring consistency with galactic dynamics leads to the value $f_{0}\simeq4.2\times10^{26}\,M_{\odot}\,\mathrm{kpc}\,(\mathrm{km/s})^{2}$ \cite{Chahboun_2024,Chahboun_2026}. This normalization can be obtained by choosing a hierarchy $\nu_{\sigma}/M_{s}\sim10^{24}$, for instance by setting $\nu_{\sigma}\simeq0.8\,\mathrm{eV}$ at matter-radiation equality ($z_c\simeq3400$). Because the ratio enters as a fourth power, even a modest hierarchy between $\nu_{\sigma}$ and $M_{s}$ produces a strongly enhanced effective coupling, making $f_{0}$ highly sensitive to the choice of the suppression scale. 

This hierarchy is most clearly interpreted through the associated Gauss-Bonnet length scale, $\ell_{\mathcal{G}}=\sqrt{16\pi G f_{0}/c^{4}}\simeq3.3\,\mathrm{kpc}$, which defines the distance below which higher-curvature effects become significant. The underlying mechanism is analogous to a seesaw \cite{Minkowski:1977sc,PhysRevD.98.015004,PhysRevD.104.075038}: in the neutrino context, a small mass results from division by a large scale, whereas in this model, a large effective coupling arises from division by a very small scale. Therefore, the model does not necessitate introducing a new heavy scale but instead depends on the existence of a very light scale that amplifies the Gauss-Bonnet interaction.

Although the coupling exhibits a large numerical value, it is not a fundamental constant of the theory. Consequently, the relative importance of the higher-curvature term is determined by the length $\ell_{\mathcal{G}}$, this length serves as an effective ultraviolet scale, marking the threshold below which the Gauss--Bonnet contribution ceases to be a perturbative correction. Its evolution throughout cosmic history depends on the field content. During inflation, with $\chi$ stabilized at the origin and $\sigma$ still confined, $f\simeq f_{\rm in}$ yields a microscopic $\ell_{\mathcal{G}}\sim10^{-30}\,\mathrm{m}$, corresponding to an energy near $10^{14}\,\mathrm{GeV}$ and comparable to the Hubble radius of that epoch. In this regime, $\ell_{\mathcal{G}}H\sim\mathcal{O}(1)$, so the higher-curvature term dominates the dynamics and drives super-inflationary behavior. After the transition in the $\sigma$ sector, the coupling attains the value quoted above and $\ell_{\mathcal{G}}$ increases to galactic scale, while the Hubble radius becomes cosmological. As a result, $\ell_{\mathcal{G}}H_{0}\sim10^{-6}$ and the Gauss--Bonnet term becomes negligible for homogeneous expansion. Thus, the same operator is horizon-scale during inflation and a short-distance correction today, with its role in each epoch determined by the vacuum expectation values of the fields to which it couples, rather than by parameter choice.

A comparison with the value required for the Gauss--Bonnet term to account for late-time acceleration is instructive. As emphasized in Ref.~\cite{Fernandes_2022}, realizing dark energy through the higher-curvature sector necessitates a coupling whose associated length is comparable to the present Hubble radius. In this scenario, $\nu_{\sigma}/M_{s}\sim 10^{27}$ and $M_{s}\simeq4.6\times10^{-37}\,\mathrm{GeV}$, resulting in $\ell_{\mathcal{G}}\simeq3.1\,\mathrm{Gpc}$ and $\ell_{\mathcal{G}}H_{0}\simeq0.7$. We do not adopt this possibility here, as the present construction attributes late-time acceleration to the inflaton on its exponential tail, with the Gauss-Bonnet coupling required only to remain dynamically negligible in the background while acting on galactic scales. This comparison highlights the strong dependence of $\ell_{\mathcal{G}}$ on the suppression scale, with six orders of magnitude in length corresponding to three in $M_{s}$. It is important to note that neither value directly impacts Solar System tests, where deviations from general relativity are constrained at the $10^{-5}$ level \cite{Bertotti:2003rm,Will:2014kxa}. A coupling with $\ell_{\mathcal{G}}$ on kiloparsec scales would, if taken at face value, modify gravity within the Solar System. The absence of such modifications indicates that the effective coupling relevant on these scales differs from the cosmological one. Therefore, some form of screening is required, as seen in the chameleon \cite{Khoury:2003aq}, symmetron \cite{Hinterbichler:2010es}, and Vainshtein \cite{Vainshtein:1972sx,Babichev:2013usa} mechanisms, where the effective coupling is suppressed in regions of high ambient density or curvature but remains unsuppressed on cosmological scales.

This behavior preserves the standard post-inflationary evolution. Given that $|\mathcal{G}|\sim24H^{4}$, the Gauss–Bonnet term's contribution to the background energy density scales as $(\ell_{\mathcal{G}}H)^{2}$, which is currently of order $10^{-13}$ and thus negligible. The effects are instead limited to perturbations on scales near $\ell_{\mathcal{G}}$. Notably, this scale corresponds to the galactic regime, where discrepancies between visible matter and observed rotation curves are observed \cite{1980ApJ...238..471R,McGaugh:2016leg}. Previous work has demonstrated that a Gauss–Bonnet coupling of this magnitude can reproduce flat rotation curves and a Tully–Fisher-like scaling \cite{Chahboun_2024}.

The subsequent evolution of $\sigma$ is governed by the curvature of the effective potential $V_{\rm eff}=V-f\mathcal{G}$ at the true vacuum \(\nu_\sigma\), to which both terms contribute,
\begin{equation}
  m_{\sigma}^{2}=2\lambda_{\sigma}\nu_{\sigma}^{2}e^{-Q\phi}-\frac{12\,\mathcal{G}\,f_{0}}{\nu_{\sigma}^{2}}\,e^{nQ\phi} .
  \label{eq71}
\end{equation}

The sign of the second term is determined by the expansion history rather than by the coupling. Given that $\mathcal{G}=24H^{2}(H^{2}+\dot H)$, the Gauss-Bonnet invariant remains negative throughout the post-inflationary era, in contrast to the quasi-de Sitter phases where $\mathcal{G}$ is positive. Consequently, both contributions to \eqref{eq71} are positive. Furthermore, the same large coupling that provides the model with its galactic length scale also enhances the stability of the vacuum responsible for its generation.

The symmetry breaking releases vacuum energy, expressed as $\Delta V=\tfrac14\lambda_{\sigma}\nu_{\sigma}^{4}e^{-Q\phi}$. It is important to specify how this energy evolves. Because $\sigma$ is insufficiently coupled to decay, the field evolves toward $\nu_{\sigma}$, overshoots, and undergoes oscillations. When $m_{\sigma}$ greatly exceeds the Hubble rate during this epoch, the oscillations remain coherent, and the average pressure over a cycle is zero. The conservation equation for \(\sigma\) is given by
\begin{equation}
  \dot\rho_{\sigma}+3H\rho_{\sigma}=0,
  \label{eq72}
\end{equation}
where \(\rho_{\sigma}(a)=\Delta V\left(a_{c}/a\right)^{3} \), as in the misalignment mechanism known from axion cosmology \cite{Preskill:1982cy,Hui:2016ltb}. The released energy is converted not into radiation but into a pressureless condensate. Although the potential vanishes at its minimum, the field does not come to rest there. Its amplitude decreases as $a^{-3/2}$ due to Hubble friction while oscillations persist, resulting in a residual $\rho_{\sigma}$ that remains today.

This contribution must be kept small, since in the present model the dark matter is to be accounted for geometrically rather than by a relic population. Expressing \eqref{eq72} today,
\begin{equation}
  \Omega_{\sigma}=\frac{T_{0}^{3}}{12M_p^2H_0^2}\nu_{\sigma}\,\lambda_{\sigma}^{-1/2}c_{\sigma}^{3/2}e^{\frac{Q\phi}{2}}.
  \label{eq73}
\end{equation}

These quantities depend on different combinations of parameters. At a fixed ratio $\lambda_{\sigma}/c_{\sigma}$, which determines the transition epoch, $\Omega_{\sigma} \propto c_{\sigma}$. Lowering both couplings simultaneously suppresses the condensate without altering the transition. For $\lambda_{\sigma}=c_{\sigma}=10^{-2}$, the same value as the
waterfall self-coupling, we obtain $\Omega_{\sigma}\simeq7\times10^{-4}$, below one per cent of the dark matter density and hence dynamically negligible, while $z_{c}\simeq3400$ is unchanged. The released energy is correspondingly a fraction $2\times10^{-3}$ of the radiation density at that time, so that no appreciable spectral distortion is induced \cite{Fixsen:1996nj}. We note that $\sigma$ must nonetheless be long-lived; a decay releasing $\Delta V$ near recombination would violate the COBE/FIRAS bounds by a wide margin, so the coupling responsible for the thermal mass must not open a rapid decay channel. This is the opposite requirement to that placed on $\chi$, which had to decay promptly in order to generate the bath at all; the two transitions share a mechanism but differ in fate, and it is the width alone that distinguishes them.

Following the $\sigma$ transition, the composition of the Universe returns to the standard configuration, with the dark sector accounted for solely by the two remaining components of the model. The waterfall condensate has already decayed into the radiation bath. The $\sigma$ condensate contributes only $\Omega_{\sigma}\lesssim10^{-3}$ and can therefore be neglected. The quintessential inflaton remains, characterized by the term $V_{0}e^{-Q\phi}$, which serves as the source of dark energy. The Friedmann equation is then given by

\begin{equation}
  3M_{p}^{2}H^{2}=\rho_{r}+\rho_{m}+\frac{1}{2}\dot\phi^{2}+V_{0}e^{-Q\phi}-24nQf_0 e^{nQ\phi}\dot{\phi}H^{3},
\end{equation}
in which $\rho_{r}\propto a^{-4}$ is the radiation inherited from the waterfall decay, $\rho_{m}\propto a^{-3}$ the baryonic matter, and the last term the higher-curvature contribution, suppressed by $(\ell_{\mathcal{G}}H)^{2}\sim10^{-13}$ and hence entirely negligible at the background level. The higher-curvature sector therefore leaves the expansion history untouched while remaining active at the scale $\ell_{\mathcal{G}}$, which is precisely the separation the construction requires. The cosmological consequences of this background, including the growth of structure and the associated dark matter phenomenology, have been analyzed in detail in Ref.~\cite{Chahboun_2026} for the case $n=1$, of which the present construction is a generalization.

\clearpage
\section{Conclusion and comments}
\label{sec:conclusion}

This work presents a cosmological framework in which a single scalar potential and a single Gauss-Bonnet coupling function, defined once and then held fixed, generate a sequence of connected epochs from a super-inflationary burst to the late-time dark sector. The distinguishing feature of this construction is not solely the inclusion of a higher-curvature term. Although couplings such as $f(\phi)\mathcal{G}$ have been extensively studied in the context of inflation \cite{Guo:2010jr,PhysRevD.88.123508,Koh:2014bka}, the key innovation is that the same operator assumes qualitatively distinct roles in each epoch, with its magnitude determined dynamically by the vacuum expectation values of the coupled fields rather than being arbitrarily specified.

The initial phase is governed by the Gauss-Bonnet term. In this regime, the coupling increases as $e^{nQ\phi}$ while the potential decreases as $e^{-Q\phi}$, resulting in the higher-curvature contribution becoming the dominant factor in the dynamics rather than a minor correction. This dynamic leads to super-inflationary expansion, characterized by $\dot H>0$. This approach contrasts with conventional models, where the Gauss-Bonnet coupling only marginally modifies slow-roll inflation \cite{Guo:2010jr}. In the present framework, the higher-curvature term dominates, enabling a closed-form solution for this phase. The interplay between the coupling and the potential also determines the termination of this phase as $\mathcal{G}\propto H^{4}$ decreases, the geometrically induced mass of the waterfall field is overtaken by its tachyonic mass, initiating instability at a field value determined solely by the combination $(n-1)Q$. This parameter combination recurs throughout the analysis, effectively reducing the parameter space to one dimension despite two couplings.

The waterfall transition supplies the radiation that the super-inflationary phase cannot generate independently. Following symmetry breaking, the resulting condensate is coherent, with $m_{\chi}\gg H$, and decays within a fraction of a Hubble time. This rapid decay promptly populates the light sector and establishes a thermal bath. As a result, this process replaces conventional reheating: no distinct epoch of inflaton decay occurs, and the Universe transitions directly from one inflationary regime to another. The mechanism resembles tachyonic preheating in hybrid models \cite{PhysRevLett.87.011601,PhysRevD.79.063531}, although the parameters in this scenario place the system firmly within the perturbative, fast-decay regime rather than the non-linear regime.

After the thermal bath forms, the inflaton dissipates into it, resulting in a warm inflationary regime \cite{PhysRevLett.75.3218,Berera:2008ar}. The model employs a two-stage configuration in which the inflaton couples exclusively to heavy mediators, yielding Boltzmann-suppressed thermal corrections to its mass and maintaining a flat potential \cite{BasteroGil:2016qru}. Two aspects of this phase are noteworthy. First, a quartic potential, previously disallowed in the cold scenario, becomes viable: the thermal enhancement of the scalar spectrum and the $Q_{\rm wi}^{-5/2}$ suppression of the tensor amplitude together reduce $r$ well below observational limits, while $n_{s}$ remains within the Planck range \cite{Planck2018X,PhysRevLett.127.151301}. Second, this outcome is robust, not fine-tuned. The observed spectral tilt persists across a range of $(n-1)Q$ values much broader than the model's own consistency window, so $n_{s}$ does not constrain the parameters. Instead, internal requirements such as the stability of the waterfall vacuum and the brief duration of the post-transition radiation era determine the parameters, with $r$ providing the primary predictive constraint.

The growth of inflaton fluctuations induced by their coupling to radiation, which would otherwise render the spectrum strongly blue with a cubic dissipation coefficient, is regulated by the shear viscosity of the thermal bath \cite{BasteroGil:2011xd}. This regulation is essential to the model's viability. The alignment of maximal damping with regions of greatest thermal enhancement is an intrinsic feature of the framework. Both phenomena depend on the ratio of the Hubble rate to the temperature, thereby reducing the sensitivity of predictions to uncertainties in the numerical fits for the growth function.

Beyond inflation, the Gauss-Bonnet coupling reemerges in an unexpected form. The third scalar, initially stabilized at the origin by the higher-curvature term and subsequently by its thermal mass, is released as the temperature drops to its symmetry-breaking scale. This transition transforms a very small mass scale into a much larger coupling, with the ratio entering as a fourth power, following logic precisely inverse to that of the seesaw mechanism \cite{Minkowski:1977sc}. The resulting coupling defines a characteristic length of a few kiloparsecs. The coincidence of this scale with the range where galactic rotation curves deviate from Newtonian predictions \cite{1980ApJ...238..471R,McGaugh:2016leg} is noteworthy. However, no definitive claim is made based solely on this order-of-magnitude argument; a thorough analysis of the perturbations is required and is reserved for future work.

Two aspects of the construction warrant emphasis. The inflaton drives both phases of inflation, reheats the Universe through its decay, and persists as the dark energy component. The waterfall field terminates super-inflation and generates the radiation bath, while the third scalar provides the large higher-curvature coupling without significantly affecting the energy budget. Thus, three scalars, one potential, and one coupling function connect the inflationary epoch to the late Universe. The higher-curvature term does not disturb the background expansion after inflation, thereby preserving the standard sequence of radiation, matter, and dark energy domination.

Several questions remain unresolved. The possibility that gravitational waves acquire a thermal enhancement in warm inflation remains debated. The dissipation coefficient and the growth function are derived from numerical fits within general relativity; a self-consistent calculation is preferable. Additionally, the model does not address the baryon asymmetry. Future research will explore these issues.

\newpage
\renewcommand\bibname{BIBLIOGRAPHY}
 \renewcommand\refname{References}

\clearpage
\bibliography{Biblio.bib}
\bibliographystyle{apsrev4-2}

\end{document}